\documentclass[reprint, aps, prx, amsmath, amssymb, superscriptaddress, longbibliography]{revtex4-2}
\usepackage{graphicx}
\usepackage{xcolor}
\usepackage{bm}
\usepackage{soul}
\usepackage[normalem]{ulem}
\usepackage[colorlinks, linkcolor= blue, citecolor = blue, urlcolor=blue]{hyperref}
\usepackage{comment}
\usepackage{physics}
\usepackage[T1]{fontenc}
\usepackage[utf8]{inputenc}
\usepackage{float}
\usepackage{pifont}
\usepackage{cancel}
\usepackage{wrapfig}
\usepackage[stable]{footmisc}
\usepackage{array}
\usepackage{gensymb}
\usepackage{multirow}
\usepackage{hhline}
\usepackage{chemformula}
\def\nn{\nonumber}
\def\bea{\begin{eqnarray}}
\def\eea{\end{eqnarray}}
\def\be{\begin{equation}}
\def\ee{\end{equation}}

\def\kb{{\bm k}}

\def\e{\varepsilon}

\def\bal{\begin{aligned}}
\def\eal{\end{aligned}}

\begin{document}
\title{Extrinsic Spin Splitter Currents in Altermagnets}
\author{Sanjay Sarkar}
\thanks{Sanjay Sarkar and Sayan Sarkar contributed equally.}
\affiliation{Department of Physics, Indian Institute of Technology Kanpur, Kanpur 208016, India}
\author{Sayan Sarkar}
\thanks{Sanjay Sarkar and Sayan Sarkar contributed equally.}
\affiliation{Department of Physics, Indian Institute of Technology Kanpur, Kanpur 208016, India}
\author{Amit agarwal}
\email{amitag@iitk.ac.in}
\affiliation{Department of Physics, Indian Institute of Technology Kanpur, Kanpur 208016, India}
\begin{abstract}
Altermagnets exhibit momentum-dependent spin splitting despite having zero net magnetization. This enables a spin-splitter effect in which an external electric field generates transverse spin currents by separating oppositely polarized carriers. Here, we develop a unified semiclassical theory of linear extrinsic spin-splitter currents, incorporating impurity-induced side-jump and skew-scattering contributions, and apply it to the $d$-wave altermagnet \ch{FeSb2}. We demonstrate that asymmetric impurity scattering provides a dominant channel for spin-splitter currents. Remarkably, the resulting extrinsic spin conductivity is time-reversal even, in contrast to previously studied spin-splitter responses arising from symmetric scattering. 
\end{abstract}
\maketitle
\section{Introduction}

Altermagnetism has recently emerged as a distinct class of magnetic order that combines key characteristics of both ferromagnets and antiferromagnets \cite{Hayami2019,smejkal_PRX2022_bey,smejkal_PRX2022_emer,Roig_prb2024,Song_NRM2025,Mostovoy_APL2025}. Altermagnets are collinear antiferromagnets with zero net magnetization, yet they exhibit pronounced momentum-dependent spin splitting of electronic bands, reminiscent of ferromagnets \cite{Yuan_prl2024,Zeng_APL2025,Yang_AFM2025,Ma_prb2025,Lee_CAP2025}. In contrast to conventional antiferromagnets, where opposite spin sublattices are related by translation or inversion, the magnetic sublattices in altermagnets are connected by crystal rotation symmetries combined with spin operations \cite{smejkal_PRX2022_bey,Turek_prb2022,jungwirth_2024,Bhowal_prx2024}. This unconventional symmetry structure leads to strongly spin-polarized itinerant electrons even in the absence of significant spin–orbit coupling (SOC) and gives rise to unconventional spin transport phenomena \cite{Karube_prl2022,Das_JPCM2023,Lyu_RP2024,kapri_prb2025,Zarzuela_prb2025,Shi_apl2025,Fu_npj2025,sarkar_2025}.

A direct consequence of momentum-dependent spin splitting in altermagnets is the spin splitter effect, in which the spin-dependent anisotropy of the electronic band structure allows an external electric field to drive carriers with opposite spin polarization in different directions, generating transverse spin currents without net charge flow~\cite{Zhang_NJP2018,zelenzy_prl2021}. Owing to its nonrelativistic origin and symmetry protection, the spin splitter effect is fundamentally distinct from conventional spin Hall responses driven by relativistic SOC~\cite{Hirsh_prl1999,Sinova_rmp2015,Zheng_prb2024,Roy_prm2022,sarkar_arXiv2025}. Recent theoretical studies have explored spin splitter currents arising from intrinsic band-structure effects as well as from symmetric impurity scattering~\cite{zelenzy_prl2021,Yi_prb2025,yang_PRL2026}. However, a systematic understanding of disorder-induced extrinsic contributions remains incomplete. In realistic materials, impurity scattering is unavoidable and can introduce additional transport channels that qualitatively modify spin responses. In particular, extrinsic mechanisms such as side-jump and skew scattering~\cite{Du2019,Datta2024,ahmed_small2025}, arising from coordinate shifts during scattering events and asymmetric scattering probabilities, have not yet been systematically investigated for spin-splitter currents in alternagnets. Understanding these contributions is crucial for establishing a complete description of spin transport in altermagnets. 

\begin{figure}
    \centering
    \includegraphics[width=1\linewidth]{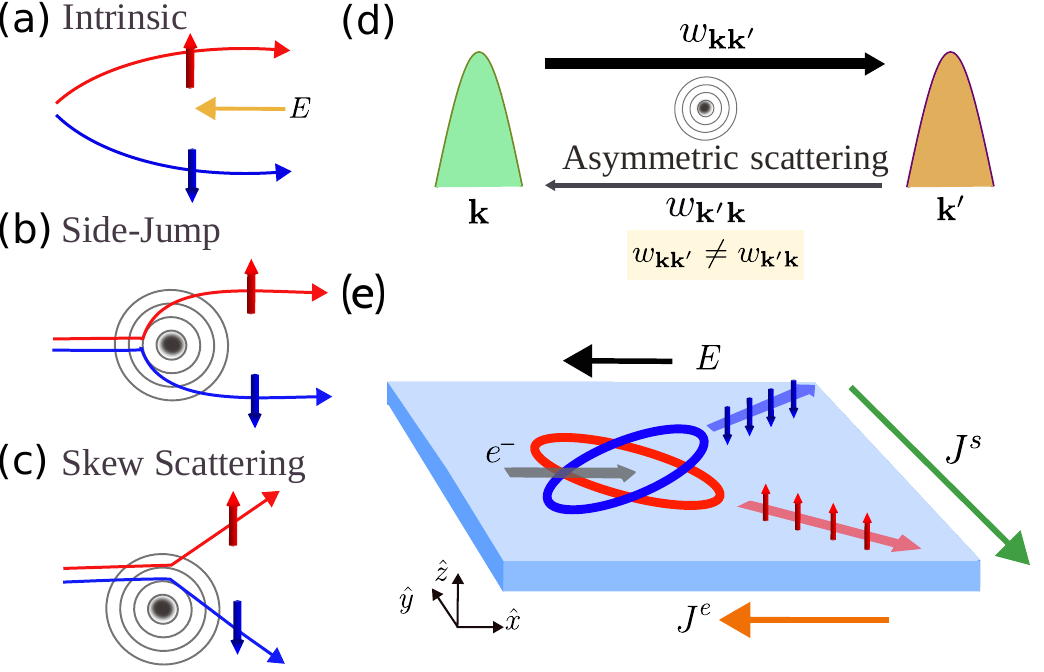}
    \caption{\textbf{Extrinsic spin-splitter current:} Schematic illustration of intrinsic and extrinsic mechanisms contributing to the spin splitter current in the presence of asymmetric impurity scattering. (a) Intrinsic mechanism: An external electric field induces an anomalous velocity, transversely deflecting opposite-spin carriers via band structure topology. (b) Side-jump mechanism: Spin-up and spin-down electrons undergo equal and opposite vertical displacements at the impurity site, preserving total momentum. (c) Skew-scattering mechanism: asymmetric impurity scattering bends spin-up and spin-down trajectories in different directions, breaking wave-vector conservation. (d) Extrinsic processes originating from asymmetric impurity scattering rates $(w_{\kb'\kb}\neq w_{\kb\kb'})$. (e) Real-space schematic of a spin-splitter current arising from the momentum-dependent spin splitting of the Fermi surface in an altermagnet.}
    \label{fig_0}
\end{figure}

Here, we address this gap by developing a semiclassical theory of the linear extrinsic spin splitter current and applying it to the $d$-wave altermagnet FeSb$_2$. We demonstrate that asymmetric impurity scattering can generate sizable spin splitter currents and, in experimentally relevant regimes, even dominate over intrinsic contributions. Remarkably, we find that these scattering induced responses are governed by band-geometric quantities, including the Berry curvature (BC) and spin Berry curvature (SBC), and are $\mathcal T$-even in nature, in contrast to earlier studies~\cite{zelenzy_prl2021}. Furthermore, we analyze the role of SOC and show that these extrinsic contributions are finite only in its presence. 


\section{Semiclassical description of spin current}
In this section, we first define the spin current operator, then derive the Boltzmann equation with symmetric and asymmetric scattering contributions and use it to obtain explicit expressions for intrinsic and extrinsic spin conductivities.

The spin current in a weakly disordered crystal subject to a weak electric field ${\bm E}$, arises from both the field-induced and disorder-induced modifications of the carrier dynamics. Within the Boltzmann transport framework, the expectation value of the spin current is expressed as, 
\begin{equation}\label{eq:cur}
J_a^{\nu} = \sum_l j_{a,l}^{\nu} f_l~,
\end{equation}
where $a$ denotes the direction of flow and $\nu$ the direction of spin polarization. The summation is over the Bloch carrier states $\ket{l}=\ket{n {\bm k}}$, with $n$ and ${\bm k}$ being the band index and crystal momentum, respectively. In Eq.~\eqref{eq:cur}, $f_l$ is the Fermi–Dirac distribution, and $j_{a,l}^{\nu}$ is the quantum-mechanical expectation value of the spin-current operator in state $l$.

The total spin-current operator, including corrections to the Bloch state induced by the external electric field~\cite{zhang_PRB2024_int} and impurity scattering~\cite{xiao_PRB2017_semi}, can be naturally decomposed as, 
\begin{equation}\label{total_cur}
j_{a,l}^{\nu} = j_{a,l}^{\nu,0} + j_{a,l}^{\nu,{\rm anm}} + j_{a,l}^{\nu,{\rm sj}}~.
\end{equation}
Here, $j_{a,l}^{\nu,0}=\langle l|\hat{j}_a^\nu|l\rangle$ is the expectation value of the spin-current operator in the unperturbed Bloch state. The spin current operator, $\hat{j}_a^{\nu}=\tfrac{1}{2}\{ \hat{s}^{\nu}, \hat{v}^a \}$, is    defined as the anticommutator of the spin ($\hat{s}^\nu$) and velocity operators ($\hat{v}^a$). 

The field-induced correction $j_{a,l}^{\nu,{\rm anm}}$ captures the anomalous contribution arising from the interband coherence in the presence of an external electric field. It is given by
\begin{equation}
j_{a,l}^{\nu,{\rm anm}}=-\frac{e}{\hbar}E_b \Omega_{l}^{\nu,ab},
\end{equation}
where 
\begin{equation}
\Omega^{\nu,ab}_{l}
=-2\hbar^{2}\, \mathrm{Im} \sum_{n' \ne n}
\frac{(j_a^\nu)_{nn'}({\bm k})\, v^b_{n'n}({\bm k})}
{(\varepsilon_{n {\bm k}} - \varepsilon_{n' {\bm k}})^{2}}~,
\label{eq:SBC}
\end{equation}
is the \textit{spin Berry curvature} (SBC). It captures the band geometric origin of the intrinsic spin Hall effect~\cite{sinova_PRL2004,guo_PRL2005,guo_PRL2008,sun_PRL2016_strong}. In Eq.~\eqref{eq:SBC},  for a given operator $\hat{A}$, we define
$A_{n'n}(\mathbf{k}) \equiv \langle u_{n' \mathbf{k}} | \hat{A} | u_{n \mathbf{k}} \rangle$,
where $|u_{n\mathbf{k}}\rangle$ denotes the cell periodic part of the unperturbed Bloch state $\ket{n\bm k}$. 

Additionally, impurity scattering modifies the spin current through a side-jump mechanism associated with coordinate shifts of itinerant spins during scattering. The extrinsic side-jump contribution $j_{a,l}^{\nu,{\rm sj}}$ represents the spin-current analogue of the side-jump velocity. It is defined as~\cite{xiao_PRB2017_semi, xiao_PRB2019_temp}
\bea\label{sjv}
j^{\nu,{\rm sj}}_{a,l}
&=& -2\pi \sum_{n' {\bm k}'} 
 W_{{\bm k}{\bm k}'} 
\delta(\varepsilon_{n {\bm k}} - \varepsilon_{n' {\bm k}'}) \nn\\
&\times&
\mathrm{Im}\Bigg[
\sum_{n'' \neq n'} 
\frac{
\langle u_{n {\bm k}} | u_{n' {\bm k}'} \rangle
\langle u_{n'' {\bm k}'} | u_{n {\bm k}} \rangle
(j_a^\nu)_{n' n''}({\bm k'})}
{\varepsilon_{n' {\bm k}'} - \varepsilon_{n'' {\bm k}'}} \nn\\
&-& 
\sum_{n'' \neq n}
\frac{
\langle u_{n'' {\bm k}} | u_{n' {\bm k}'} \rangle
\langle u_{n' {\bm k}'} | u_{n {\bm k}} \rangle
(j_a^\nu)_{n n''}({\bm k})}
{\varepsilon_{n {\bm k}} - \varepsilon_{n'' {\bm k}}}
\Bigg].
\eea
Here, \( W_{{\bm k},{\bm k}'} = W_{{\bm k}',{\bm k}} \) 
is the Born amplitude describing impurity scattering. 
For static impurities, \( W_{{\bm k},{\bm k}'} = \langle |V^{0}_{{\bm k},{\bm k}'}|^{2} \rangle_{\rm dis} \), 
with \( \langle \cdots \rangle_{\rm dis} \) denoting configurational averaging over random impurity distributions. The detailed derivation of Eq.~\eqref{sjv} is provided in Appendix~\ref{app_D}. 

\subsection{Semiclassical Boltzmann equation}\label{SBE}

\begin{figure*}
    \centering
    \includegraphics[width=1\linewidth]{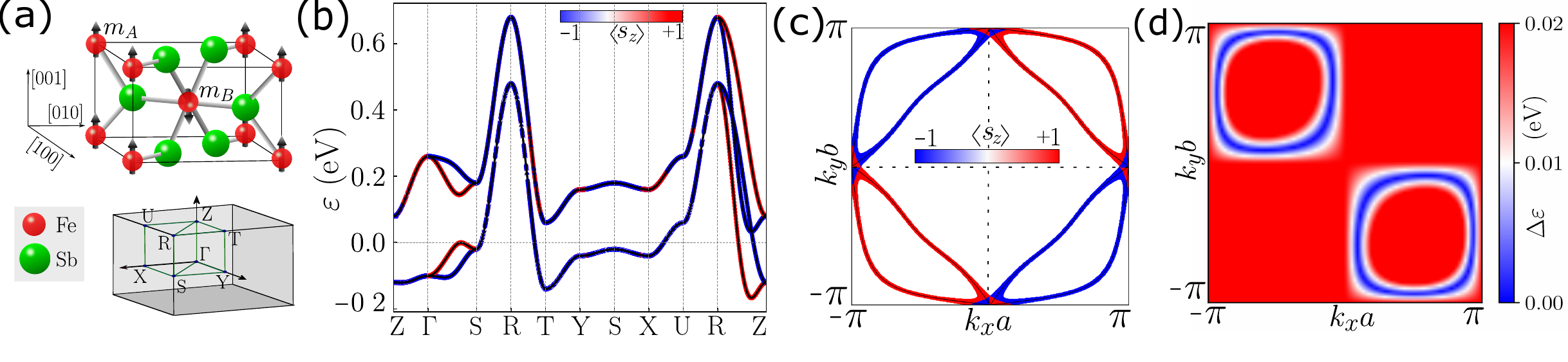}
    \caption{{\bf Spin-split electronic structure of doped \ch{FeSb2}:} (a) Crystal structure and Brillouin zone (BZ) of \ch{FeSb2}. (b) Spin-projected band-structure along the high symmetry path $\rm Z$-$\rm\Gamma$-$\rm S$-$\rm R$-$\rm T$-$\rm Y$-$\rm S$-$\rm X$-$\rm U$-$\rm R$-$\rm Z$, showing pronounced non-relativistic spin splitting along the $\rm \Gamma$-$\rm S$ and $\rm Z$-$\rm R$ directions. (c) Spin-projected Fermi surface on the $k_x-k_y$ plane at the chemical potential $\mu=0.16$ eV, exhibiting the characteristic $d$-wave anisotropy of the spin-split bands. (d) The
band splitting, $\Delta\varepsilon$, between the two lowest spin-splitted bands on the $k_x-k_y$ plane within the BZ.}
    \label{fig_2}
\end{figure*}

To describe disorder-induced spin transport, we employ the semiclassical Boltzmann transport formalism. In the presence of elastic impurity scattering and homogeneous perturbations, the non-equilibrium occupation function $f_l$ satisfies~\cite{ashcroft_book1976_solid}
\begin{equation}
\frac{\partial f_l}{\partial t} + \dot{\bm{k}}\!\cdot\!\partial_{\bm{k}} f_l = I_{\text{el}}\{ f_l \}~.
\end{equation}
Here, $I_{\text{el}}\{f_l\}$ is the elastic collision integral accounting for disorder induced scattering between Bloch states. Using the Born approximation, it is given by 
\begin{equation}\label{col_int}
I_{\text{el}}\{f_l\} = -\sum_{l'} (w_{l'l} f_l - w_{ll'} f_{l'})~.
\end{equation}
The scattering rate $w_{ll'}$ between states $l$ and $l'$ is governed by Fermi’s Golden rule~\cite{ashcroft_book1976_solid}, 
\begin{equation}
w_{ll'} =\frac{2\pi}{\hbar}  
\Big\langle 
\big| \langle l | V_{\text{imp}} | {l'}^{\rm dis} \rangle \big|^{2}
\Big\rangle_{\text{dis}} 
\delta(\varepsilon_l - \varepsilon_{l'})~.
\end{equation}
Here, $V_{\mathrm{imp}}$ represents the impurity potential, and $|{l'}^{\rm dis}\rangle$ is the eigenstate of the full Hamiltonian $H(\bm{k}) = H_{0}(\bm{k}) + V_{\mathrm{imp}}$. It  satisfies the Lippmann-Schwinger equation~\cite{sakurai2017QM}, 
\begin{equation}\label{eq:LippmannSchwinger}
|l^{\rm dis}\rangle = |l\rangle 
+\frac{V_{\text{imp}}}{\varepsilon_l - H_0 + i\eta} ~|l^{\rm dis}\rangle~,
\end{equation}
where the infinitesimal $i\eta\to0^+$ ensures the outgoing boundary condition for the scattering states. 

Although microscopic time-reversal symmetry enforces constraints on scattering amplitudes, disorder averaging and higher-order processes can generate antisymmetric components in the effective scattering rate. Thus, in general $w_{ll'} \neq w_{l'l}$. 
It is therefore convenient to decompose the scattering rate into its symmetric and antisymmetric parts as
\begin{equation}
w_{ll'}^{\mathrm{S}} = w_{l'l}^{\mathrm{S}} = \frac{w_{ll'} + w_{l'l}}{2}~,~~
w_{ll'}^{\mathrm{A}} = -w_{l'l}^{\mathrm{A}} = \frac{w_{ll'} - w_{l'l}}{2}~.
\end{equation}
The symmetric part $w_{ll'}^{\mathrm{S}}$ describes conventional relaxation processes, 
typically treated within the relaxation-time approximation. 
In contrast, the antisymmetric component $w_{ll'}^{\mathrm{A}}$ gives rise to skew-scattering contributions.

For weak disorder, the scattering rate can be systematically expanded in powers of the impurity potential. Retaining terms up to fourth order in the impurity potential~\cite{xiao_PRB2019}, we have 
\begin{equation}\label{w_order}
w_{l'l} \approx w_{l'l}^{(2)} + w_{l'l}^{(3),A} + w_{l'l}^{(4),A}~.
\end{equation}
Here, $w_{l'l}^{(3), A}$ and $w_{l'l}^{(4),A}$ denote the antisymmetric contributions arising at third and fourth order, respectively. These terms give rise to skew-scattering effects. The symmetric parts of these higher-order terms merely renormalize the leading second-order contribution and are therefore neglected.

The second-order term can be further decomposed as
\begin{equation}\label{w_2}
w_{l'l}^{(2)} = w_{l'l}^{(\rm 2S)} + w_{l'l}^{(2),\mathrm{cs}}~,
\end{equation}
with the symmetric field-independent part
\begin{equation}
w_{l'l}^{(\rm 2S)} = \frac{2\pi}{\hbar} \langle |V_{l'l}|^{2} \rangle_{\rm dis}\, 
\delta(\varepsilon_l - \varepsilon_{l'})~,
\end{equation}
and 
\begin{equation}
w_{l'l}^{(2),\mathrm{cs}} = \frac{2\pi }{\hbar} 
\langle |V_{l'l}|^{2} \rangle_{\rm dis}\,
\frac{\partial \delta(\varepsilon_l - \varepsilon_{l'})}{\partial \varepsilon_l}\,
e{\bm E}\cdot\delta {\bm r}_{l'l}~,
\end{equation}
captures the coordinate-shift correction induced by the external electric field  $\bm{E}$. This term originates from the real-space displacement of the electron wave packet during an impurity scattering event. The coordinate shift $\delta \bm{r}_{l'l}$ of the wave packet is given by~\cite{sinitsyn_PRB2006}
\begin{equation}\label{coordinate}
\delta {\bm r}_{l'l} = 
\langle u_{l'} | i\partial_{\bm{k}'} | u_{l'} \rangle
-\langle u_l | i\partial_{\bm{k}} | u_l \rangle
-(\partial_{\bm{k}}+\partial_{\bm{k}'})\arg(V_{l'l})~.
\end{equation}
Here, the operator ``$\arg$'' denotes the phase (argument) of a complex number. This coordinate shift plays a crucial role in the side-jump contribution to spin transport.  Having categorized the scattering rates by symmetry and physical origin, we next decompose the collision integral and the nonequilibrium distribution function into intrinsic, side-jump, and skew-scattering parts.

\subsection{Decomposition of the collision integral and distribution function}

\begin{figure*}
    \centering
    \includegraphics[width=0.8\linewidth]{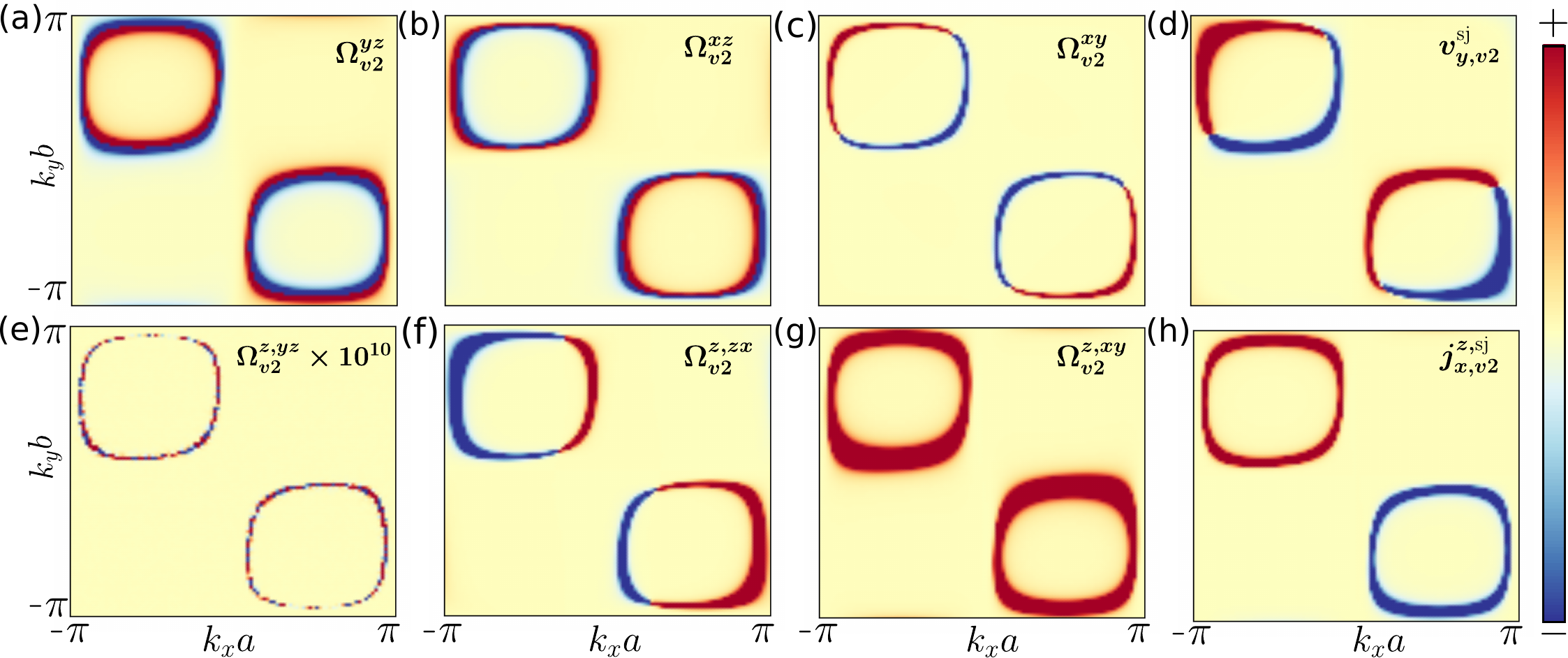}
    \caption{{\bf Band-geometric quantities relevant for extrinsic spin-splitter transport in \ch{FeSb2}:} (a-c) Momentum-space distribution of Berry curvature components $\Omega^{yz}_{v2}$, $\Omega^{zx}_{v2}$, and $\Omega^{xy}_{v2}$
, and (d) the corresponding side-jump velocity component 
$v_{y,v2}^{\rm sj}$
, evaluated for the lowest valence band ($v2$) on the 
$k_x-k_y$
 plane within the BZ. (e-g) Momentum-space distribution of Spin Berry curvature components $\Omega^{z,yz}_{v2}$, $\Omega^{z,zx}_{v2}$, and $\Omega^{z,xy}_{v2}$
, and (h) the spin-dependent side-jump velocity 
$j_{x,v2}^{z,\rm sj}$ for the same band.
}
    \label{fig_3}
\end{figure*}

Building on the different physical origins of the impurity scattering rates discussed above, the associated elastic collision integral can be naturally decomposed into distinct contributions. In particular, we separate the collision term into intrinsic, side-jump, and skew-scattering components, $I_{\mathrm{el}}(f_l) = I_{\mathrm{el}}^{\mathrm{in}}(f_l) + I_{\mathrm{el}}^{\mathrm{sj}}(f_l) + I_{\mathrm{el}}^{\mathrm{sk}}(f_l),$
and these distinct components have the form~\cite{Du2019}

\begin{subequations}
\label{eq:collision_terms}
\begin{align}
I_{\mathrm{el}}^{\mathrm{in}}(f_l) &= - \sum_{l'} w_{ll'}^{\mathrm{2S}} (f_l - f_{l'})~, \label{eq:inelastic_term} \\[4pt]
I_{\mathrm{el}}^{\mathrm{sj}}(f_l) &= -\sum_{l'} w_{ll'}^{(2),\mathrm{cs}} (f_l - f_{l'})~, \label{eq:sj_term} \\[4pt]
I_{\mathrm{el}}^{\mathrm{sk}}(f_l) &= \sum_{l'} w_{ll'}^{\mathrm{A}} (f_l + f_{l'})~. \label{eq:sk_term}
\end{align}
\end{subequations}
Using the antisymmetry property 
$w_{ll'}^{\mathrm{A}} = - w_{l'l}^{\mathrm{A}}$, 
the skew-scattering collision term reduces to a symmetric combination of distribution functions, 
$(f_l + f_{l'})$.  The intrinsic term 
$I_{\mathrm{el}}^{\mathrm{in}}$
 originates from symmetric second-order scattering and governs conventional momentum relaxation. The side-jump contribution 
$I_{\mathrm{el}}^{\mathrm{sj}}$
 arises from the coordinate-shift correction to scattering in the presence of an external electric field, while the skew-scattering term 
$I_{\mathrm{el}}^{\mathrm{sk}}$
 reflects the antisymmetric part of the scattering rate. As discussed in Section~\ref{SBE}, the antisymmetric scattering rate 
$w_{ll'}^{\mathrm{A}}$ originates from higher-order impurity scattering and can be further decomposed into third- and fourth-order contributions [see Eq.~\eqref{w_order}]. 

Accordingly, the skew-scattering collision integral naturally splits into two physically distinct channels~\cite{guo_PRB2024_ext},
\begin{equation}
I_{\mathrm{el}}^{\mathrm{sk}}(f_l)
=
I_{\mathrm{el}}^{\mathrm{sk3}}(f_l)
+
I_{\mathrm{el}}^{\mathrm{sk4}}(f_l)~,
\end{equation}
with
\begin{subequations}
\begin{align}
I_{\mathrm{el}}^{\mathrm{sk3}}(f_l)
&=
\sum_{l'}
w_{ll'}^{(3),A}(f_l+f_{l'})~,\\
I_{\mathrm{el}}^{\mathrm{sk4}}(f_l)
&=
\sum_{l'}
w_{ll'}^{(4),A}(f_l+f_{l'})~.
\end{align}
\end{subequations}
Here, the term 
$I_{\mathrm{el}}^{\mathrm{sk3}}$ describes the non-Gaussian skew-scattering contribution arising at third order in the impurity potential. In contrast,  
$I_{\mathrm{el}}^{\mathrm{sk4}}$ represents the Gaussian skew scattering contribution originating from fourth-order scattering processes.

Combining all these terms, the nonequilibrium distribution function can be expanded as, 
\begin{equation}\label{eq:f_decomposition}
f_l = f_l^{\mathrm{in}} + f_l^{\mathrm{sj}} + f_l^{\mathrm{sk3}}+ f_l^{\mathrm{sk4}}~.
\end{equation}
Here, $f_l^{\mathrm{in}}$ represents the intrinsic part determined by symmetric scattering, while the remaining terms encode disorder-induced corrections linear in ${\bm E}$. Considering the steady state distribution function with $\frac{\partial f_l}{\partial  t}=0$ and substituting Eq.~\eqref{eq:f_decomposition} into the Boltzmann equation yields the coupled relations
\begin{subequations}\label{BZ_couple_eq}
\begin{align}
\dot{\bm{k}}\cdot\partial_{\bm{k}} f^{\mathrm{in}}_l
&= I^{\mathrm{in}}_{\mathrm{el}}(f^{\mathrm{in}}_l)~, \\
\dot{\bm{k}}\cdot\partial_{\bm{k}} f^{\mathrm{sj}}_l
&= I^{\mathrm{in}}_{\mathrm{el}}(f^{\mathrm{sj}}_l) + I^{\mathrm{sj}}_{\mathrm{el}}(f^{\mathrm{in}}_l)~, \\
\dot{\bm{k}}\cdot\partial_{\bm{k}} f^{\mathrm{sk3}}_l
&= I^{\mathrm{in}}_{\mathrm{el}}(f^{\mathrm{sk3}}_l) + I^{\mathrm{sk3}}_{\mathrm{el}}(f^{\mathrm{in}}_l)~, \\
\dot{\bm{k}}\cdot\partial_{\bm{k}} f^{\mathrm{sk4}}_l
&= I^{\mathrm{in}}_{\mathrm{el}}(f^{\mathrm{sk4}}_l) + I^{\mathrm{sk4}}_{\mathrm{el}}(f^{\mathrm{in}}_l)~.
\end{align}
\end{subequations}
Here, we have neglected the mixed terms between side-jump and skew-scattering channels as their effects are subleading in the weak-disorder limit. We solve these coupled equations perturbatively in Appendix~\ref{app_A} to obtain linear-order corrections to $f_l$, and use them to determine the intrinsic and extrinsic contributions to the spin splitter current. With the nonequilibrium distribution function determined for each scattering channel, we are now in a position to evaluate the corresponding contributions to the linear spin current and spin conductivity.

\subsection{Linear spin current and spin conductivity}

We now evaluate each linear spin conductivity contribution explicitly and identify its scaling with the relaxation time and its symmetry character. Expanding the general expression for the spin current [Eq.~\eqref{eq:cur}] to first order in ${\bm E}$, we obtain
\bea
j_a^{\nu} &=& \sum_l \Big[
j_{a,l}^{\nu,0}
\big(f_l^{\mathrm{in},(1)} + f_l^{\mathrm{sj},(1)} + f_l^{\mathrm{sk3},(1)}+ f_l^{\mathrm{sk4},(1)}\big) \nn\\
&& +~ j_{a,l}^{\nu,{\rm anm}} f_l^0
+ j_{a,l}^{\nu,{\rm sj}} f_l^{{\rm in},(1)}
\Big]~.
\eea
Thus, the total spin current can be decomposed into additive contributions associated with distinct microscopic mechanisms,
\begin{equation}
j_a^{\nu}
= j_a^{\nu,{\rm in}}
+ j_a^{\nu,{\rm sj}}
+ j_a^{\nu,{\rm sk3}}
+ j_a^{\nu,{\rm sk4}}~.
\end{equation}
The individual components are given by, 
\begin{subequations}\label{eq:j_total}
\begin{align}
j_{a}^{\nu,{\rm in}} &= \sum_l 
\left( j_{a,l}^{\nu,0} f_l^{\rm in,(1)} + j_{a,l}^{\nu,{\rm anm}} f_l^0 \right)~, \\
j_a^{\nu,{\rm sj}} &= \sum_l 
\left( j_{a,l}^{\nu,0} f_l^{\rm sj,(1)} + j_{a,l}^{\nu,{\rm sj}} f_l^{{\rm in},(1)} \right)~,\\
j_a^{\nu,{\rm sk3}} &= \sum_l j_{a,l}^{\nu,0} f_l^{\rm sk3,(1)}, \\
j_a^{\nu,{\rm sk4}} &= \sum_l j_{a,l}^{\nu,0} f_l^{\rm sk4,(1)}~
\end{align}
\end{subequations}
\begin{table}[t!]
    \centering
    \caption{Symmetry transformation of momentum-dependent physical quantities under parity ($\mathcal{P}$) and time-reversal symmetry ($\mathcal{T}$). The symbols $\cancel{\mathcal P}$ and $\cancel{\mathcal T}$ indicate broken parity and time-reversal symmetry, respectively.}
    {
    \begin{tabular}{c@{\hskip 0.8cm} c@{\hskip 0.8cm} c}
    \hline \hline
	\rule{0pt}{3ex}
    Quantities & $ {\mathcal P},~\cancel{\mathcal T}$ & ${\mathcal T},~\cancel{\mathcal P}$
    \\[2ex]
    \hline \hline
	\rule{0pt}{3ex}
    $\bm{k}$ & $-{\bm k}$ & $-{\bm k}$ 
    \\[2ex]
    
    $\varepsilon_n({\bm k})$ & $\varepsilon_n({-\bm k})$ & $\varepsilon_n(-{\bm k})$
    \\[2ex]
   
    $v_{a,n}({\bm k})$ & $-v_{a,n}(-{\bm k})$ & $-v_{a,n}(-{\bm k})$
    \\[2ex]

    $j^{\nu,0}_{a,n}({\bm k})$ & $-j^{\nu,0}_{a,n}(-{\bm k})$ & $j^{\nu,0}_{a,n}(-{\bm k})$
    \\[2ex]
    
    $\Omega^{ab}_n({\bm k})$ & $\Omega^{ab}_n(-{\bm k})$ & $-\Omega^{ab}_n(-{\bm k})$
    \\[2ex]
    
    $\Omega^{\nu,ab}_n({\bm k})$ & $\Omega^{\nu,ab}_n(-{\bm k})$ & $\Omega^{\nu,ab}_n(-{\bm k})$
    \\[2ex]
  
    $v_{a,n}^{\rm sj}({\bm k})$ & -$v_{a,n}^{\rm sj}(-{\bm k})$ & $v_{a,n}^{\rm sj}(-{\bm k})$
    \\[2ex]
  
    $j_{a,n}^{\nu,\rm sj}({\bm k})$ & $-j_{a,n}^{\nu,\rm sj}(-{\bm k})$ & $-j_{a,n}^{\nu,\rm sj}(-{\bm k})$
    \\[2ex]
    $ w^{(3),A}_n(\bm k, \bm k')$ & $ w^{(3),A}_n(-\bm k, -\bm k')$ & $ -w^{(3),A}_n(-\bm k, -\bm k')$ \\[2ex]
    $ w^{(4),A}_n(\bm k, \bm k')$ & $ w^{(4),A}_n(-\bm k, -\bm k')$ & $ -w^{(4),A}_n(-\bm k, -\bm k')$ \\[2ex]
    \hline \hline
			
\end{tabular}}
 \label{table_geo_quant_symmetry}
\end{table}
These terms correspond to intrinsic, side-jump, non-Gaussian conventional skew scattering, and Gaussian skew scattering mechanisms, respectively. The corresponding spin conductivity tensor, $\sigma_{ab}^\nu$, defined through $j_a^\nu = \sigma_{ab}^\nu E_b$, can therefore be written as\begin{equation}
\sigma_{ab}^{\nu}=
\sigma_{ab}^{\nu,{\rm in}}
+
\sigma_{ab}^{\nu,{\rm sj}}
+
\sigma_{ab}^{\nu,{\rm sk3}}
+
\sigma_{ab}^{\nu,{\rm sk4}}~.
\end{equation}
Explicitly, the individual contributions are given by, 
\begin{subequations}\label{eq:sigma_total}
\begin{align}
\sigma_{ab}^{\nu,{\rm in}} 
&= \frac{e}{\hbar}\sum_l 
\left( 
j_{a,l}^{\nu,0} \tau \partial_{k_b} f_l^{(0)} 
- \Omega_l^{\nu,ab} f_l^{(0)}
\right)~, \\[2pt]
\sigma_{ab}^{\nu,{\rm sj~}} 
&= \frac{e\tau}{\hbar} \sum_l 
\left( 
j_{a,l}^{\nu,{\rm sj}} \partial_{k_b} f_l^{(0)} 
- j_{a,l}^{\nu,0} \hbar v_{b,l}^{{\rm sj}} 
\frac{\partial f_l^{(0)}}{\partial \varepsilon_l}
\right)~, \\[2pt]
\sigma_{ab}^{\nu,{\rm sk3}} 
&= \frac{e\tau^2}{\hbar} \sum_{l,l'}w^{(3),A}_{ll'}
\left( 
j_{a,l}^{\nu,0} - j_{a,l'}^{\nu,0}
\right) 
\partial_{k_b} f_l^{(0)}~, \\[2pt]
\sigma_{ab}^{\nu,{\rm sk4}} 
&= \frac{e\tau^2}{\hbar} \sum_{l,l'}w^{(4),A}_{ll'}
\left( 
j_{a,l}^{\nu,0} - j_{a,l'}^{\nu,0}
\right) 
\partial_{k_b} f_l^{(0)}~.
\end{align}
\end{subequations}
Here, $\tau$ denotes the transport relaxation time arising from the symmetric second-order scattering rate $w^{(2S)}$. We emphasize that the intrinsic Drude term and the side-jump contribution scale linearly with $\tau$, whereas the skew-scattering terms scale as $\tau^{2}$, reflecting their higher-order origin in asymmetric impurity potential.

To explicitly calculate the extrinsic spin conductivities, we consider a minimal disorder model consisting of short-range, randomly distributed $\delta$-function impurities. We have, 
\be 
V(\bm{r})=\sum_i V_i\delta(\bm{r}-\bm{R}_i)~,
\ee
where $V_i$ is the impurity strength at position ${\bm R}_i$, and the summation runs over all impurity sites. Following the systematic simplifications developed in Refs.~\cite{varshney2026} and detailed in Appendix~\ref{app_B}, \ref{app_C}, and \ref{app_D}, the side-jump velocity and the antisymmetric scattering rates can be expressed directly in terms of the Berry curvature and SBC. We obtain, 
\begin{subequations}
 \begin{align}
     {\bm v}^{\rm sj}_{n, \kb} &= \frac{2\pi}{\hbar} n_i V_0^2 \mathcal{D}_n(\e_{n {\bm k}})  [ \kb \times {\bm \Omega}_n(\bm k)] ~,\label{v_sj_comp} \\ 
      {\bm j}^{\nu,\rm sj}_{n, \kb} &= \frac{2\pi}{\hbar} n_i V_0^2 \mathcal{D}_n(\e_{n {\bm k}})  [ \kb \times {\bm \Omega}^\nu_n(\bm k)] ~,\label{j_sj_comp}
      \\
     w^{(3), A}_{n, \kb\kb' } &= \frac{2\pi^2}{\hbar} n_i V_1^3 \mathcal{D}_n(\e_{n {\bm k}}) [(\kb \times \kb') \cdot {\bm \Omega}_n(\bm k)]\nonumber \\
     &\quad \times \delta(\e_{n {\bm k}}  - \e_{n {\bm k'}})~,\label{w_3a_comp} \\ 
     w^{(4), A}_{n, \kb\kb' } &= \frac{2\pi^2}{\hbar} n^2_i V_0^4 \mathcal{D}_n(\e_{n {\bm k}}) [(\kb \times \kb') \cdot \tilde{\bm \Omega}_n(\bm k)]\nonumber \\
     &\quad \times \delta(\e_{n {\bm k}}  - \e_{n {\bm k'}})~.\label{w_4a_comp}
 \end{align}   
\end{subequations}
Here, $\tilde{\bm \Omega}_n(\bm{k})=\sum_{n'}{\bm \Omega}_{nn'}(\bm{k})/(\varepsilon_{n\bm{k}}-\varepsilon_{n'\bm{k}})$ denotes the energy-normalized Berry curvature, and $\mathcal{D}_n(\e_{n {\bm k}})$ is the density of states of band $n$ at energy $\varepsilon_{n\bm{k}}$. 
In deriving these simplified expressions, we assume that either inversion symmetry ($\mathcal P$) or time-reversal symmetry ($\mathcal T$) enforces $\varepsilon_{n}(\bm{k})=\varepsilon_{n}(-\bm{k})$. 
 These results  highlight that all extrinsic spin-transport contributions are governed by band-geometric quantities in nonmagnetic or centrosymmetric materials. 

The symmetry properties summarized in Table~\ref{table_geo_quant_symmetry} allow us to determine the time-reversal character of the individual spin-conductivity contributions in Eq.~\eqref{eq:sigma_total}. The Drude-like intrinsic term in $\sigma^{\nu, \rm in}_{ab}$ is odd under $\cal{T}$~\cite{sarkar_arXiv2025}. In contrast, the remaining contributions, namely the anomalous term, the side-jump term, and the skew-scattering spin-conductivity contributions remain invariant under $\cal{T}$. 
This $\mathcal{T}$-even character of the anomalous term originates from the transformation properties of the spin current operator, and should be contrasted with the anomalous charge Hall response, which is $\mathcal{T}$-odd. As a result, the extrinsic spin-splitter response predicted here has finite $\cal{T}$-even contribution, in sharp contrast to previously discussed spin-splitter currents that are $\cal{T}$-odd~\cite{zelenzy_prl2021} in presence of symmetric scattering.

\section{Extrinsic Spin-Splitter current in $\mathrm{FeSb_2}$} 

In this section, we apply the semiclassical framework developed above to a realistic altermagnetic material, \ch{FeSb2}. Our goal is to quantify the relative magnitude of intrinsic and extrinsic spin-splitter currents for realistic systems. The momentum-dependent $d$-wave spin splitting of \ch{FeSb2} makes it a natural platform to investigate intrinsic and extrinsic spin-splitter currents. \ch{FeSb2} belongs to the centrosymmetric space group $Pnnm$. Figure~\ref{fig_2}(a) shows the crystal structure and Brillouin zone (BZ) of \ch{FeSb2}. The lattice parameters of the crystal are given as $a=5.83$ \AA, $b=6.54$ \AA~and $c=3.18$ \AA~\cite{Petrovic_prb2005,Zhang_CPB2025}. We employ a minimal tight-binding model that captures the essential symmetry properties and electronic structure of this spin-split altermagnetic system. We use the model parameters from Ref.~\cite{Roig_prb2024}. 

The effective tight-binding Hamiltonian of \ch{FeSb2} is given by,  
\begin{figure}[t!]
    \centering
    \includegraphics[width=0.75\linewidth]{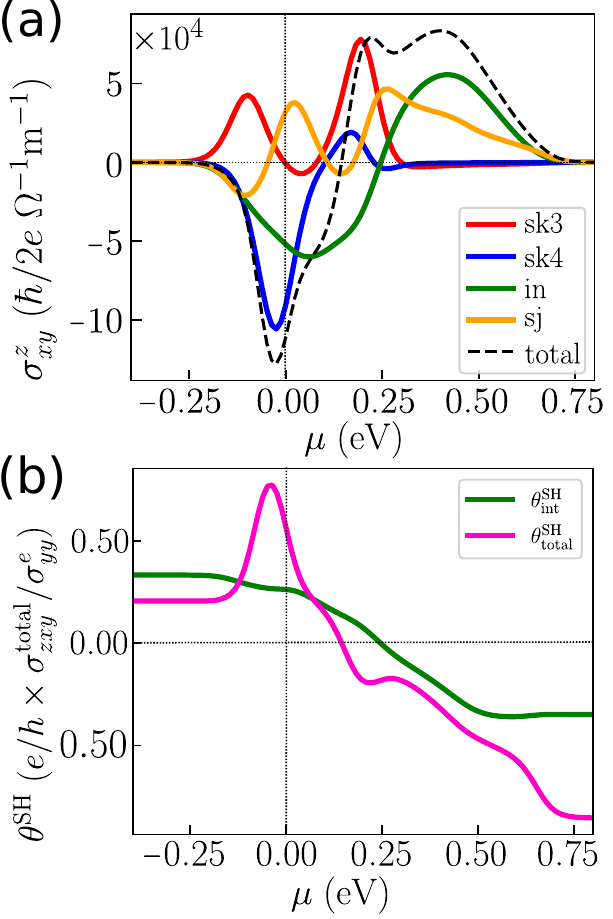}
    \caption{{\bf Extrinsic transverse spin conductivity and spin Hall angle in altermagnetic \ch{FeSb2}:}  (a) Chemical-potential ($\mu$) dependence of different  $\sigma_{xy}^z$ components, showing  contributions from sk3, sk4, side-jump (sj), intrinsic mechanisms, and their total sum. (b) The corresponding spin Hall angle $\theta^{\rm SH}=\sigma^{z,\rm total}_{xy}/\sigma^e_{yy}$ as a function of $\mu$. Here, $\sigma^e_{yy}$ denotes the longitudinal charge conductivity. The spin Hall angle reaches values as large as $\theta^{\rm SH}_{\rm total}\sim 0.8$, significantly exceeding the intrinsic contribution. This indicates highly efficient spin–charge conversion driven by extrinsic scattering mechanisms in altermagnetic \ch{FeSb2}.}
    \label{fig_4}
\end{figure}

\be\label{Ham_AM}
{\cal H} = \varepsilon_{0,{\bm k}} + t_{x,{\bm k}} \tau_x + t_{z,{\bm k}} \tau_z + \tau_y \vec{\lambda}_{{\bm k}} \cdot \vec{\sigma} + \tau_z \vec{J} \cdot \vec{\sigma}~.
\ee
Here, $\tau_i$ and $\sigma_i$ are the Pauli matrices in the orbital (sublattice) and spin space, respectively. The term $\varepsilon_{0,\bm k}$ represents the sublattice-independent dispersion, while 
$t_{x,\bm k}$ and $t_{z,\bm k}$ describe inter- and intra-sublattice hopping processes. The vector $\vec{\lambda}_{{\bm k}}$ encodes the intrinsic spin–orbit coupling (SOC), and $\vec{J}$ denotes the N\'eel order parameter.
For the given space group $Pnnm$, the momentum-dependent coefficients entering the Hamiltonian take the form
\begin{equation}\label{eq:model_params}
\begin{aligned}
\varepsilon_{0,\bm k} & = -\mu+
t_{1x}\cos (k_xa) + t_{1y}\cos (k_yb) + t_2\cos (k_zc)
  \\
& + t_3\cos (k_xa) \cos (k_yb) + t_{4x}\cos (k_xa) \cos (k_zc)
 \\
& + t_{4y}\cos (k_yb) \cos (k_zc)+t_5\cos (k_xa) \cos (k_yb) \cos (k_zc) ~, \\
t_{x,\bm k} & = t_8 \cos\frac{k_xa}{2}\cos\frac{k_yb}{2}\cos\frac{k_zc}{2}~, \\
t_{z,\bm k} & = t_6 \sin (k_xa) \sin (k_yb) + t_7 \sin (k_xa) \sin (k_yb) \cos (k_zc)~, \\
\lambda_{x,\bm k} & = \lambda_{x0} \sin\frac{k_xa}{2}\cos\frac{k_yb}{2}\sin\frac{k_zc}{2}~, \\
\lambda_{y,\bm k} & = \lambda_{y0} \cos\frac{k_xa}{2}\sin\frac{k_yb}{2}\sin\frac{k_zc}{2}~,\nn\\
\lambda_{z,\bm k} & = \lambda_{z0} \cos\frac{k_xa}{2}\cos\frac{k_yb}{2}\cos\frac{k_zc}{2}~. 
\end{aligned}
\end{equation}
%
For \ch{FeSb2}, we use the parameter values $t_{1x}=-0.1,t_{1y}=-0.05,t_2=-0.05, t_3=0.06, t_{4x}=0.1,t_{4y}=0.05,t_5=-0.05,t_6=0.05, t_7=-0.1, t_8=0.15$ and $J_z=0.1$ (all are in eV). We also include finite SOC strengths $\lambda_{x0}=\lambda_{y0} = 10~\mathrm{meV}$ and $\lambda_{z0} = 0$. 

The spin-projected band structure is displayed in Fig.~\ref{fig_2}(b), where a clear momentum-dependent spin splitting is visible along the $\Gamma$--$\rm S$ and $\rm R$--$\rm Z$ path. The corresponding spin-resolved Fermi surface in the $k_x$-$k_y$ plane at chemical potential $\mu=0.16~\rm eV$ is shown in Fig.~\ref{fig_2}(c), highlighting the characteristic $d$-wave form of the spin splitting. Fig.~\ref{fig_2}(d) shows the energy difference $\Delta\varepsilon = \varepsilon_{\uparrow} - \varepsilon_{\downarrow}$, between the two lowest spin-split bands on the $k_x$-$k_y$ plane within the BZ. The blue regions, forming two diagonal square-like shapes, mark the region in the vicinity of the band touching point where band-geometric effects are strongly enhanced. Consequently, the Berry curvature, spin Berry curvature, side-jump velocity, and spin side-jump velocity shown in Fig.~\ref{fig_3} are also peaked near these regions.
\begin{figure}[t!]
    \centering
    \includegraphics[width=0.75\linewidth]{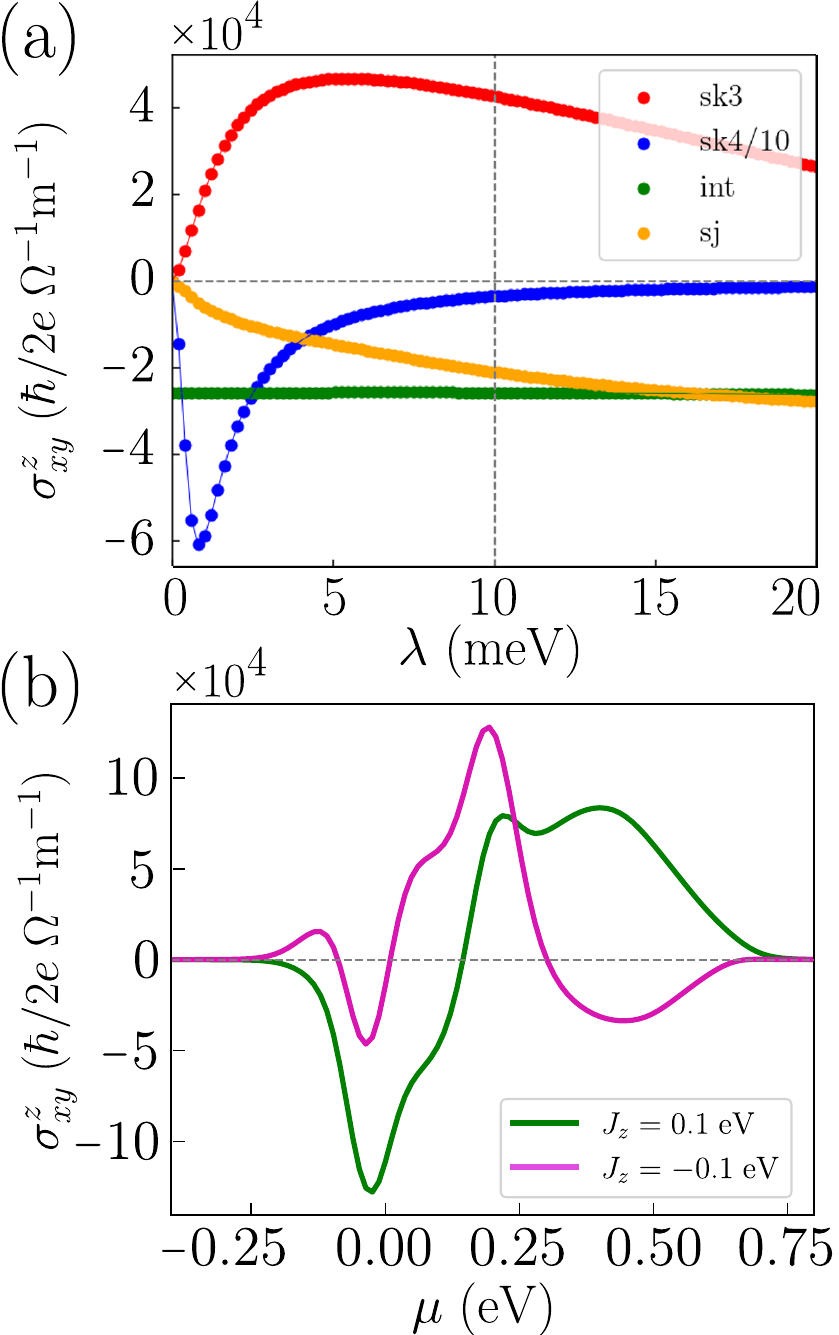}
    \caption{{\bf Role of spin–orbit coupling and N\'eel order in extrinsic spin-splitter current:} (a) The variation of different extrinsic contributions to the transverse spin conductivity with SOC strength in doped-\ch{FeSb2} with chemical potential $\mu=-0.1~\rm eV$, (b) Chemical-potential dependence of the transverse spin conductivity for opposite orientations of the N\'eel order, parametrized by $J_z$ parameter of the Hamiltonian. These highlight the sensitivity of the spin splitter current to SOC strength and N\'eel vector direction.}
    \label{fig_5}
\end{figure}

The resulting extrinsic spin conductivity components, obtained by integrating these quantities over the BZ, are presented as functions of chemical potential in Fig.~\ref{fig_4}(a). For the calculation of skew-scattering conductivities, we have assumed the impurity density to be $10^{19}~\rm cm^{-3}$ and the impurity potential moments to be $V_0=4.94\times10^{2}~\rm eV.$\AA$^3$ and $V_1=3.4\times10^1~\rm eV$.\AA$^2$ (see Appendix B for details). As shown in Fig.~\ref{fig_4}(a), the extrinsic spin-splitter currents can be comparable in magnitude to both intrinsic and symmetric-scattering contributions. 
These results establish that asymmetric impurity scattering plays a key role in determining the spin-splitter response in altermagnets. Similar dominance of asymmetric scattering has been reported in charge transport phenomena such as the anomalous Hall and nonlinear anomalous Hall effects.

We further quantify the efficiency of spin–charge conversion by evaluating the spin Hall angle 
$\theta^{\mathrm{SH}}$. It is defined as the ratio of the transverse spin Hall current to the longitudinal charge current, $\theta^{\mathrm{SH}}=(e/\hbar)J^s_{zx}/J^e_{y}=(e/\hbar)\sigma^{z}_{xy}/\sigma^e_{yy}$. Figure~\ref{fig_4}(b) shows the variation of 
$\theta^{\mathrm{SH}}$ with chemical potential 
$\mu$. The results reveal a significant increase in the spin Hall angle upon inclusion of extrinsic spin-current contributions. Notably, the spin Hall angle reaches values as large as 
$\theta^{\mathrm{SH}}\sim 0.8$, indicating a highly efficient spin-current generation. This places \ch{FeSb2} among the most efficient spin–charge converters reported in altermagnetic systems~\cite{Chen_CPL2025, Wei_prb2025}. Importantly, this large spin-charge conversion efficiency arises primarily from extrinsic asymmetric scattering. 

\subsection{SOC and N\'eel vector dependence}
Beyond doping, spin–orbit coupling (SOC) plays a crucial role in determining the magnitude of the extrinsic spin-splitter current. Figure~\ref{fig_5}(a) shows the dependence of various spin-conductivity components on the SOC strength. We find that both the side-jump and skew-scattering contributions vanish in the absence of SOC, whereas the symmetric-scattering (Drude-type) contribution remains finite even when SOC is zero. This behavior follows from the explicit dependence of the skew-scattering and side-jump terms on band-geometric quantities such as the Berry curvature and spin Berry curvature. For the altermagnetic model (Eq.~\ref{Ham_AM}) considered here, the Berry curvature components vary linearly with the SOC strength ($\lambda$), depending on the spin component of SOC parallel to the N\'eel order. Consequently, in the absence of SOC, the Berry curvature vanishes identically. Moreover, a finite spin Berry curvature requires interband spin mixing, which is also absent when SOC is zero. Therefore, SOC is essential for enabling the asymmetric impurity-scattering processes that generate the extrinsic spin-splitter current.

\begin{figure}
    \centering
    \includegraphics[width=0.75\linewidth]{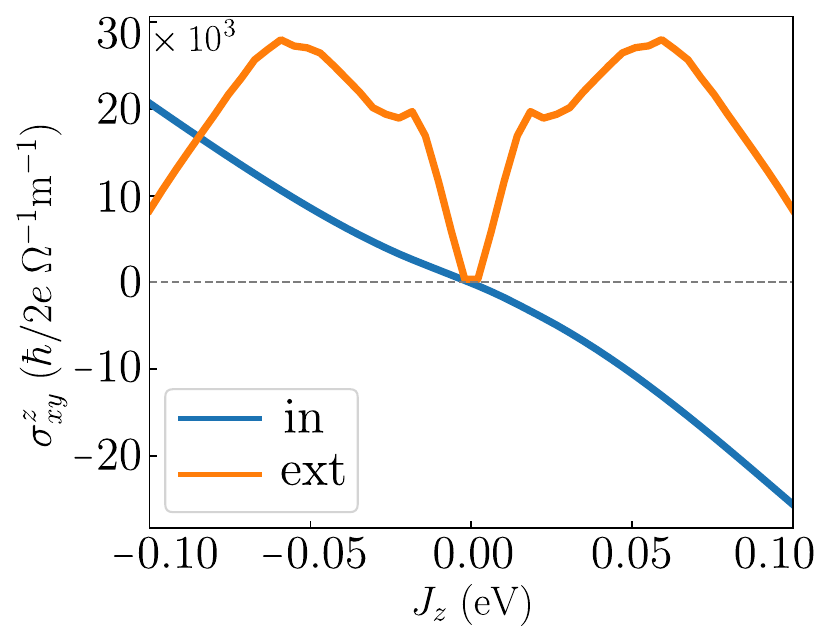}
    \caption{\textbf{Variation of conductivity with the magnetic order:} The figure shows how the extrinsic and intrinsic conductivity vary with the altermagnetic order parameter $J_z$. The intrinsic component (Drude), being a $\cal T$-odd quantity, changes sign while $J_z$ goes from positive to negative value. The extrinsic component doesn't change sign with $J_z$ and can be written as a series of even powers of $J_z$.}
    \label{fig_6}
\end{figure}

We next investigate the effect of reversing the N\'eel vector by changing the sign of $J_z$ in the Hamiltonian. The resulting spin conductivity as a function of chemical potential is shown in Fig.~\ref{fig_5}(b). Notably, the spin-splitter current does not simply reverse sign under N\'eel vector reversal, indicating the presence of a finite $\mathcal{T}$-even contribution. This behavior contrasts with earlier expectations for spin-splitter currents in altermagnets~\cite{zelenzy_prl2021,Yi_prb2025,Jeong_AS2025} and highlights the subtle symmetry properties of extrinsic spin transport in these systems.

The $\mathcal{T}$-even nature of the extrinsic spin conductivity has important physical implications. As shown explicitly in Fig.~\ref{fig_6}, the extrinsic (side-jump and skew-scattering) contributions depend only on even powers of the altermagnetic order parameter $J_z$. This implies that these components can yield finite responses proportional to $\mathcal{N}^2$, where $\mathcal{N}$ denotes the N\'eel order. Such spin currents generate damping-like spin–orbit torques~\cite{Farokhnezhad_IOP2023}. In altermagnetic systems, damping-like torques can drive THz-frequency oscillations of the N\'eel vector by inducing a canting between oppositely aligned magnetic sublattices~\cite{Oleksandr2024,Khymyn_SR2017,Tomasz_prb2017}.

\section{Conclusion}
In summary, we developed a unified semiclassical theory of the linear extrinsic spin-splitter current in altermagnets, systematically incorporating disorder-induced side-jump and skew-scattering mechanisms. Going beyond previous studies focused on intrinsic and symmetric-scattering contributions~\cite{zelenzy_prl2021,Yi_prb2025}, our results demonstrate that contributions arising from asymmetric scattering can dominate the spin-splitter current. Applying the theory to the $d$-wave altermagnet \ch{FeSb2}, we showed that asymmetric impurity scattering generates sizable transverse spin currents that can exceed intrinsic contributions, yielding a large spin Hall angle ($\sim 0.8$) and highly efficient spin–charge conversion.

Our symmetry analysis reveals that while the conventional Drude contribution is $\cal{T}$-odd, the extrinsic spin-splitter conductivities are intrinsically $\cal{T}$-even. Our numerical results for \ch{FeSb2} also reflect this and the spin conductivity does not simply reverse sign under N\'eel-vector reversal. We further showed that these contributions require finite relativistic spin–orbit coupling. Taken together, our findings establish disorder induced extrinsic scattering as a constructive mechanism for generating and controlling spin currents in altermagnets, providing new directions for altermagnetic spintronic functionalities.

\section{Acknowledgment}
We acknowledge many fruitful discussions with Harsh Varshney (IIT Kanpur). Sanjay Sarkar thanks the MHRD, India, for funding through the Prime Minister’s Research Fellowship (PMRF). Sayan Sarkar acknowledges IIT Kanpur for funding support.  A. A. acknowledges funding from the Core Research Grant by Anusandhan National Research Foundation (ANRF, Sanction No. CRG/2023/007003), the Department of Science and Technology of the Government of India, for Project No. DST/NM/TUE/QM-6/2019(G)-IIT Kanpur. Department of Science and Technology, India.  A. A.  acknowledges the high-performance computing facility at IIT Kanpur, including HPC 2013, and Param Sanganak.

\onecolumngrid
\appendix

\section{Calculation of the nonequilibrium distribution function from the Boltzmann equation}\label{app_A}
In this Appendix, we solve Eq.~\eqref{BZ_couple_eq} explicitly in order to
extract the individual contributions to the nonequilibrium distribution
function. In the presence of a uniform electric field,
$\dot{\bm k} = -e\bm E/\hbar$, the coupled Boltzmann equations become
\begin{subequations}\label{BZ_coupl_eq:2}
\begin{align}
-\frac{e\bm E}{\hbar}\!\cdot\!\partial_{\bm k} f_l^{\rm in}
&= I_{\rm el}^{\rm in}(f_l^{\rm in})\label{BZ_eq1}~, \\
-\frac{e\bm E}{\hbar}\!\cdot\!\partial_{\bm k} f_l^{\rm sj}
&= I_{\rm el}^{\rm in}(f_l^{\rm sj}) + I_{\rm el}^{\rm sj}(f_l^{\rm in})\label{BZ_eq2}~, \\
-\frac{e\bm E}{\hbar}\!\cdot\!\partial_{\bm k} f_l^{\rm sk3}
&= I_{\rm el}^{\rm in}(f_l^{\rm sk3}) + I_{\rm el}^{\rm sk3}(f_l^{\rm in})\label{BZ_eq3}~, \\
-\frac{e\bm E}{\hbar}\!\cdot\!\partial_{\bm k} f_l^{\rm sk4}
&= I_{\rm el}^{\rm in}(f_l^{\rm sk4}) + I_{\rm el}^{\rm sk4}(f_l^{\rm in})\label{BZ_eq4}~.
\end{align}
\end{subequations}
We treat the intrinsic elastic collision integral within the
relaxation-time approximation, introducing a characteristic
scattering time $\tau$ such that
\begin{equation}
I_{\rm el}^{\rm in}(f_l^{\rm in})
=
-\frac{f_l^{\rm in} - f_l^0}{\tau},
\end{equation}
Now, to solve Eq.~\eqref{BZ_coupl_eq:2} for the distribution functions perturbatively, we expand the non-equilibrium distribution function in powers of the electric field, 
\begin{equation}
f_l = f^{(0)}_l + f_l^{(1)} + f_l^{(2)} + \cdots,
\end{equation}
where $f_l^{(i)} \propto |{\bm E}|^{i}$ for all $i > 0$,  denotes the $i$th order correction. The intrinsic linear-response solution then follows directly as
\begin{equation}
f_l^{\rm in,(1)}
=
\frac{e\tau}{\hbar}\,
\bm E\!\cdot\!\partial_{\bm k} f_l^0.
\end{equation}
Solving Eq.~\eqref{BZ_eq2} to linear order in $\bm E$ yields the side-jump
correction
\begin{equation}
f_l^{\rm sj,(1)}
=
\tau \sum_{l'} w^{(2),{\rm cs}}_{l'l}
\bigl(f_l^0 - f_{l'}^0\bigr).
\end{equation}
Using
\(
v_l^a\,\partial_{\varepsilon_l}\delta(\varepsilon_l-\varepsilon_{l'})
=
\hbar^{-1}\partial_{k_a}\delta(\varepsilon_l-\varepsilon_{l'})
\)
and integrating by parts yields the compact form
\begin{equation}
f_l^{\rm sj,(1)}
=
- e\tau\,
\bm E\!\cdot\!\bm v_l^{\rm sj}
\,
\frac{\partial f^0}{\partial \varepsilon_l},
\end{equation}
where the side-jump velocity is
\begin{equation}\label{sj_vel}
\bm v_l^{\rm sj}
=
-\sum_{l'} w^{(2S)}_{l'l}\,
\boldsymbol{\delta r}_{ll'}.
\end{equation}
Proceeding analogously, the third- and fourth-order antisymmetric scattering
processes generate
\begin{align}
f_l^{\rm sk3,(1)}
&=
-\frac{e\tau^2}{\hbar}
\sum_{l'} w^{(3),A}_{l'l}
\left(
\bm E\!\cdot\!\partial_{\bm k} f_l^0
+
\bm E\!\cdot\!\partial_{\bm k} f_{l'}^0
\right), \\
f_l^{\rm sk4,(1)}
&=
-\frac{e\tau^2}{\hbar}
\sum_{l'} w^{(4),A}_{l'l}
\left(
\bm E\!\cdot\!\partial_{\bm k} f_l^0
+
\bm E\!\cdot\!\partial_{\bm k} f_{l'}^0
\right).
\end{align}
These results provide the linear-order nonequilibrium distribution functions
entering the evaluation of the extrinsic spin-splitter current in the main text.

\section{Simplification of scattering rates}\label{app_B}
In this section, we derive simplified expressions for the impurity scattering rates entering the semiclassical Boltzmann equation. Our goal is to reduce the general scattering amplitudes to forms suitable for numerical evaluation of the spin conductivity.
In the presence of elastic impurity scattering, the nonequilibrium distribution function $f_l$ satisfies the semiclassical Boltzmann equation
\begin{equation}
\frac{\partial f_l}{\partial t} + \dot{\bm{k}}\!\cdot\!\partial_{\bm{k}} f_l = I_{\text{el}}\{ f_l \}~,
\end{equation}
where $I_{\text{el}}\{f_l\}$ is the elastic collision integral. Within the Born approximation, it can be written as
\begin{equation}\label{col_int}
I_{\text{el}}\{f_l\} = -\sum_{l'} (w_{l'l} f_l - w_{ll'} f_{l'})~,
\end{equation}
where the scattering rate $w_{l'l}$ between states $l$ and $l'$ is governed by Fermi’s Golden rule,
\be\label{ScRt}
w_{ll'} =\frac{2\pi}{\hbar}  
\left \langle 
|\bra{l} V \ket{l^{\rm dis}}|^{2}
\right \rangle_{\text{dis}} 
\delta(\varepsilon_l - \varepsilon_{l'}).
\ee
Expanding the Lippmann–Schwinger equation up to second order in the impurity potential yields
\bea
\ket{l^{\rm dis}}&=&\ket{l}+(\varepsilon_l-H_0+i\eta)^{-1}(\hat{V}+\hat{V}(\varepsilon_l-H_0+i\eta)\hat{V})\ket{l}\nn\\&=&\ket{l}+\sum_{l'}\frac{V_{l'l}}{(\varepsilon_l-\varepsilon_{l'}+i\eta)}\ket{l'}+\sum_{l'l''}\frac{V_{l'l''}V_{l''l}}{(\varepsilon_l-\varepsilon_{l'}+i\eta)(\varepsilon_l-\varepsilon_{l''}+i\eta)}\ket{l'}~.\nn
\eea
Substituting this expansion into Eq.~\eqref{ScRt}, we get
\be
   w_{ll'}=w_{ll'}^{(2)}+w_{ll'}^{(3)}+w_{ll'}^{(4)}+\cdots,
\ee
where $w_{ll'}^{(\nu)}\propto V^\nu$ represent the $\nu$-th order scattering, for $\nu=2,3,4\cdots$. The explicit expressions of these scattering rates are given by

\bea
\label{scrt2}
w_{ll'}^{(2)}&=&\frac{2\pi}{\hbar}  
\left \langle 
V_{ll'}V_{l'l}
\right \rangle_{\text{dis}} 
\delta(\varepsilon_l - \varepsilon_{l'})~,\\
\label{scat3}
w_{ll'}^{(3)}&=&\frac{2\pi}{\hbar}  
\sum_{l''}\bigg[\frac{\langle 
V_{ll''}V_{l''l'}V_{l'l}
\rangle_{\text{dis}}}{\varepsilon_{l'} - \varepsilon_{l''}+i\eta} +\frac{\langle 
V_{ll'}V_{l'l''}V_{l''l}
\rangle_{\text{dis}}}{\varepsilon_{l'} - \varepsilon_{l''}-i\eta}\bigg]\delta(\varepsilon_l - \varepsilon_{l'})~,\\
\label{scat4}
w_{ll'}^{(4)}&=&\frac{2\pi}{\hbar}  
\sum_{l''l'''}\bigg[\frac{\langle 
V_{ll'''}V_{l'''l'}V_{l'l''}V_{l''l}
\rangle_{\text{dis}}}{(\varepsilon_{l'} - \varepsilon_{l''}-i\eta)(\varepsilon_{l'} - \varepsilon_{l'''}+i\eta)}+\frac{\langle 
V_{ll''}V_{l''l'''}V_{l'''l'}V_{l'l}
\rangle_{\text{dis}}}{(\varepsilon_{l'} - \varepsilon_{l''}+i\eta)(\varepsilon_{l'} - \varepsilon_{l'''}+i\eta)}\nn\\&&+\frac{\langle 
V_{ll'}V_{l'l'''}V_{l'''l''}V_{l''l}
\rangle_{\text{dis}}}{(\varepsilon_{l'} - \varepsilon_{l''}-i\eta)(\varepsilon_{l'} - \varepsilon_{l'''}-i\eta)}\bigg]\delta(\varepsilon_l - \varepsilon_{l'})~. 
\eea

The second-order scattering rate $w_{ll'}^{(2)}$ is strictly symmetric under exchange of initial and final states, $l\leftrightarrow l'$. On the other hand $w_{ll'}^{(3)}$ and $w_{ll'}^{(4)}$ don't follow such relations. So one can actually separate the symmetric and anti-symmetric parts of these scattering rates as follows,
\bea
    w_{ll'}^{(\nu),S}=\frac{w_{ll'}^{(\nu)}+w_{l'l}^{(\nu)}}{2}~,~~~w_{ll'}^{(\nu),A}=\frac{w_{ll'}^{(\nu)}-w_{l'l}^{(\nu)}}{2}~.
\eea
The symmetric parts of these higher-order terms merely renormalize the leading second-order contribution and do not generate qualitatively new transport phenomena. Consequently, for the third- and fourth-order scattering processes, only their antisymmetric components are relevant for transport. With this separation, the elastic collision integral naturally decomposes into symmetric and antisymmetric parts as,$I^s_{\text{el}}\{f_l\} = -\sum_{l'} w_{l'l}^{(2)} (f_l - f_{l'})$ and $I^a_{\text{el}}\{f_l\} = -\sum_{l'} (w_{l'l}^{(3),A}+w_{l'l}^{(4),A} )(f_l + f_{l'})$, respectively.

If we take into account the work done by the electric field as an electron gets displaced within the unit cell during
the collision, the second-order in the impurity part, is modified as
\bea
\label{scrt2}
w_{ll'}^{(2)}=\frac{2\pi}{\hbar}  
\left \langle 
V_{ll'}V_{l'l}
\right \rangle_{\text{dis}} 
\delta(\varepsilon_l - \varepsilon_{l'}+e{\bm E}\cdot \delta {\bm r}_{ll'})~,
\eea
where $\delta {\bm r}_{ll'}$ is the coordinate shift defined in Eq.~\eqref{coordinate} in the main text. Here, we only consider the modification to the second-order part and neglect mixed contributions from skew scattering and coordinate shifts, as they are subleading in the weak-disorder limit. Hence, the second-order part is decomposed as a symmetric field-independent part and a coordinate shift correction term induced by the external electric field [see Eq.~\eqref{w_2}].

The symmetric field-independent part governs conventional momentum relaxation, the coordinate shift correction term gives the side jump contribution, and the antisymmetric part encodes asymmetric scattering processes responsible for skew-scattering contributions to the spin response. Our next step is to simplify these scattering rates and express them in forms suitable for numerical evaluation of the spin conductivity. We begin with the symmetric second-order scattering rate $w_{ll'}^{(2S)}$,
\bea
w_{ll'}^{(2S)}&=&\frac{2\pi}{\hbar}  
\left \langle 
V_{ll'}V_{l'l}
\right \rangle_{\text{dis}} 
\delta(\varepsilon_l - \varepsilon_{l'})=\frac{2\pi}{\hbar}  
\left \langle 
V_{\kb\kb'}V_{\kb'\kb}
\right \rangle_{\text{dis}} \langle u_{ n \bm k}|u_{ n' \bm k'}\rangle \langle u_{ n' \bm k'}|u_{ n \bm k}\rangle
\delta(\varepsilon_{ n \bm k} - \varepsilon_{ n'\bm k'})~.
\eea
Here we have used the notation $V_{ll'}=\langle l|V|l'\rangle=\langle n \bm{k}|V| n'\kb'\rangle$. To simplify the expressions above, we adopt a simplest model for disorder consisting of short-range, randomly distributed $\delta$-function scatterers, $V(\bm{r})=\sum_i V_i\delta(\bm{r}-\bm{R}_i)$. Within this model, the impurity matrix element between Bloch states takes the form
\bea
   \bra{ n \bm k}V\ket{ n'\bm k'}&=&\sum_i\int d{\bm r}V_i\delta(\bm{r}-\bm{R}_i) e^{i(\bm{k}'-\bm{k})\cdot \bm{r}} \langle u_{ n \bm k}|u_{ n' \bm k'}\rangle = \sum_i V_i e^{i(\bm{k}'-\bm{k})\cdot \bm{R}_i} \langle u_{ n \bm k}|u_{ n' \bm k'}\rangle= V^0_{\kb\kb'}\langle u_{ n \bm k}|u_{ n' \bm k'}\rangle~.\nn
\eea
and the disorder average can be further simplified to
$$\left \langle 
V_{\kb\kb'}V_{\kb'\kb}\right \rangle_{\text{dis}}=\left\langle\sum_{ij}V_i V_j \exp\{(\kb'-\kb)(R_i-Rj)\}\right\rangle_{\rm dis}=n_iV_0^2~,$$
where $n_i$ is the impurity concentration and $V_0$ is the zeroth moment of the impurity potential, with units of energy times volume. We additionally take the assumption that scattering only happens within the same band but different momentum i.e $n'=n$. This condition finally simplifies the symmetric scattering rate to be
\be\label{ScRt2}
w_{n,\kb\kb'}^{(2S)}=\frac{2\pi n_i V_0^2}{\hbar} \langle u_{ n \bm k}|u_{ n \bm k'}\rangle \langle u_{ n \bm k'}|u_{ n \bm k}\rangle\delta(\varepsilon_{ n \bm k} - \varepsilon_{ n \bm k'})~.
\ee

We now turn to the antisymmetric third-order scattering rate, $$w^{(3),A}=\frac{1}{2}(w_{ll'}^{(3)}-w_{l'l}^{(3)})~.$$ Substituting Eq.~\eqref{scat3} into this definition, we get 
\bea\label{sct3}
w_{ll'}^{(3),A}&=&(\pi/\hbar)  
\sum_{l''}\bigg[\frac{\langle 
V_{ll''}V_{l''l'}V_{l'l}
\rangle_{\text{dis}}}{\varepsilon_{l} - \varepsilon_{l''}+i\eta} 
+\frac{\langle 
V_{ll'}V_{l'l''}V_{l''l}
\rangle_{\text{dis}}}{\varepsilon_{l} - \varepsilon_{l''}-i\eta}
-\frac{\langle 
V_{l'l''}V_{l''l}V_{ll'}
\rangle_{\text{dis}}}{\varepsilon_{l} - \varepsilon_{l''}+i\eta} 
-\frac{\langle 
V_{l'l}V_{ll''}V_{l''l'}
\rangle_{\text{dis}}}{\varepsilon_{l} - \varepsilon_{l''}-i\eta}\bigg]\delta(\varepsilon_l - \varepsilon_{l'})\nn\\
&=& (\pi/\hbar) 
\sum_{l''}\bigg[\langle 
V_{ll''}V_{l''l'}V_{l'l}
\rangle_{\rm dis}\left (\frac{1}{\varepsilon_{l} - \varepsilon_{l''}+i\eta}-\frac{1}{\varepsilon_{l} - \varepsilon_{l''}-i\eta} \right)
-cc.\bigg]\delta(\varepsilon_l - \varepsilon_{l'})\nn\\
&=&(4\pi^2/\hbar){\rm Im}\sum_{l''}\langle V_{ll''}V_{l''l'}V_{l'l}
\rangle_{\rm dis}\delta(\varepsilon_{l} - \varepsilon_{l'})\delta(\varepsilon_{l} - \varepsilon_{l''})\nn\\
&=& (4\pi^2/\hbar){\rm Im}\sum_{n''} \sum_{ \bm k''}\langle V_{\kb\kb''}^0 V_{\kb''\kb'}^0 V_{\kb'\kb}^0
\rangle_{\rm dis}\langle u_{ n \bm k}|u_{ n'' \bm k''} \rangle\langle u_{ n'' \bm k''}|u_{ n' \bm k'}\rangle \langle u_{ n' \bm k'}|u_{ n \bm k}\rangle\delta(\varepsilon_{ n \bm k} - \varepsilon_{ n' \bm k'})\delta(\varepsilon_{ n\bm k} - \varepsilon_{ n'' \bm k''})~.\nn\\
\eea
To further simplify the above expression, we adopt the same set of approximations used previously. In particular, we assume that the Fermi surface resides within a single band and that impurity scattering occurs only between states belonging to the same band index 
$ n$ but with different crystal momenta 
$\bm k$~\cite{guo_PRB2024_ext}. Under this intraband-scattering assumption, the intermediate band indices are constrained to satisfy $ n''= n'= n$. With all these restrictions, we first evaluate the disorder-averaged third-order impurity correlator as 
$$\langle V_{\kb\kb''}^0 V_{\kb''\kb'}^0 V_{\kb'\kb}^0
\rangle_{\rm dis}=\sum_{i,j,k}V_i V_j V_k \exp\left((\kb''-\bm{k})R_i+(\kb'-\kb'')R_j+(\bm{k}-\kb')R_k\right)=n_iV_1^3~,$$
where $n_i$ is the impurity concentration and $V_1$ is the first moment of the impurity potential, with units of $energy\times volume^{2/3}$. Finally the antisymmetric third-order scattering rate (non-Gaussian skew scattering) takes the compact form
\bea
w_{n,\kb\kb'}^{(3),A}=\frac{4\pi^2 n_i V_1^3}{\hbar}{\rm Im}\sum_{\bm k''}\langle u_{ n \bm k}|u_{ n \bm k''} \rangle\langle u_{ n \bm k''}|u_{ n \bm k'}\rangle \langle u_{ n \bm k'}|u_{ n \bm k}\rangle\delta(\varepsilon_{ n \bm k} - \varepsilon_{ n \bm k'})\delta(\varepsilon_{ n\bm k} - \varepsilon_{ n \bm k''})~.
\eea
In a similar manner, we calculate $w^{(4),A}_{ll'}$. It is given by, 
\be
   w^{(4),A}_{ll'}=-\dfrac{4\pi^2}{\hbar}\sum_{l''l'''}{\rm Im}\left [\langle V_{ll'''}V_{l'''l'}V_{l'l''}V_{l''l}\rangle_{\rm dis}-\langle V_{ll''}V_{l''l'''}V_{l'''l'}V_{l'l}\rangle_{\rm dis}-\langle V_{ll'''}V_{l'''l''}V_{l''l'}V_{l'l}\rangle_{\rm dis}\right ]\dfrac{\delta(\varepsilon_l-\varepsilon_{l'})\delta(\varepsilon_l'-\varepsilon_{l''})}{(\varepsilon_{l'}-\varepsilon_{l'''})}~.
\ee
The structure of this scattering rate differs from the previous cases. In particular, a purely intraband scattering approximation, as adopted earlier, is insufficient to yield a finite contribution. This will become clearer in the latter part of this section. To proceed, we therefore assume $l, l', l'' = (n,\bm{k}), (n,\bm{k'}), (n,\bm{k''})$, while $l''' = (n',\bm{k'''})$, where $n'$ is allowed to differ from $n$. Under this assumption, the expression simplifies to
\bea\label{sct4}
   w^{(4),A}_{n,\kb\kb'}&=&-\dfrac{4\pi^2}{\hbar}\sum_{ n'}\sum_{\kb''\kb'''}{\rm Im}\bigg[\langle V_{\kb\kb'''}V_{\kb'''\kb'}V_{\kb'\kb''}V_{\kb''\kb}\rangle_{\rm dis}\langle u_{ n\bm{k}} | u_{ n'\kb'''} \rangle \langle u_{ n'\kb'''} | u_{ n\kb'} \rangle \langle u_{ n\kb'} | u_{ n\kb''} \rangle
   \langle u_{ n\kb''} | u_{ n\bm{k}} \rangle \nn\\
   & & -\langle V_{\kb\kb''}V_{\kb''\kb'''}V_{\kb'''\kb'}V_{\kb'\kb}\rangle_{\rm dis}\langle u_{ n\bm{k}} | u_{ n\kb''} \rangle \langle u_{ n\kb''} | u_{ n'\kb'''} \rangle \langle u_{ n'\kb'''} | u_{ n\kb'} \rangle
   \langle u_{ n\kb'} | u_{ n\bm{k}}\rangle \nn\\
   & & -\langle V_{\kb\kb'''}V_{\bm{k'''k''}}V_{\kb''\kb'}V_{\kb'\kb}\rangle_{\rm dis}\langle u_{ n\bm{k}} | u_{ n'\kb'''} \rangle \langle u_{ n'\kb'''} | u_{ n\kb''} \rangle \langle u_{ n\kb''} |u_{ n\kb'} \rangle
   \langle u_{ n\kb'} | u_{ n\bm{k}}\rangle\bigg]\dfrac{\delta(\varepsilon_{ n \bm{k}}-\varepsilon_{\bm{ n k'}})\delta(\varepsilon_{ n\kb'}-\varepsilon_{ n\kb''})}{(\varepsilon_{ n\kb'}-\varepsilon_{ n'\kb'''})}\nn\\
   &=&-\dfrac{4\pi^2 n_i^2 V_0^4}{\hbar}\sum_{ n'}\sum_{\kb''\kb'''}{\rm Im}\bigg[\langle u_{ n\bm{k}} | u_{ n'\kb'''} \rangle \langle u_{ n'\kb'''} | u_{ n\kb'} \rangle \langle u_{ n\kb'} | u_{ n\kb''} \rangle
   \langle u_{ n\kb''} | u_{ n\bm{k}} \rangle -\langle u_{ n\bm{k}} | u_{ n\kb''} \rangle \langle u_{ n\kb''} | u_{ n'\kb'''} \rangle \times\nn\\ & &\langle u_{ n'\kb'''} | u_{ n\kb'} \rangle
   \langle u_{ n\kb'} | u_{ n\bm{k}}\rangle -\langle u_{ n\bm{k}} | u_{ n'\kb'''} \rangle \langle u_{ n'\kb'''} | u_{ n\kb''} \rangle \langle u_{ n\kb''} |u_{ n\kb'} \rangle
   \langle u_{ n\kb'} | u_{ n\bm{k}}\rangle\bigg]\dfrac{\delta(\varepsilon_{ n \bm{k}}-\varepsilon_{\bm{ n k'}})\delta(\varepsilon_{ n\kb'}-\varepsilon_{ n\kb''})}{(\varepsilon_{ n\kb'}-\varepsilon_{ n'\kb'''})}~. \nn \\ 
\eea
Here we have applied Wick's theorem to factorize the disorder average as 
$\langle V_{\kb\kb''} V_{\kb''\kb'} V_{\kb'\kb'''} V_{\kb'''\kb} \rangle_{\rm dis}
=
\langle V_{\kb\kb''} V_{\kb''\kb'} \rangle_{\rm dis}
\langle V_{\kb'\kb'''} V_{\kb'''\kb} \rangle_{\rm dis}.$ 
To proceed further, we need to explicitly evaluate the Bloch-state overlaps entering the disorder matrix elements.

Due to the short-range scattering potential, the electron experiences a small momentum transfer during scattering events. This leads to $q=(\kb'-\bm{k})\to 0$. Within this limit, we expand the Bloch state $\ket{u_{ n \kb'}}$ around $\bm{k}$ as
\be\label{expand_u}
\ket{u_{ n \kb'}}=\ket{u_{ n \bm{k}}}+q_b \ket{\partial_b u_{ n \bm{k}}}+\frac{1}{2}q_b q_c\ket{\partial_b \partial_c u_{ n \bm{k}}}+\cdots~.
\ee
Here we have taken Einstein's summation condition for coordinate indices $b$. This leads to
\be\label{expand_uu}
\langle u_{ n \bm k}|u_{ n \bm k'}\rangle\approx 1-iq_b\mathcal{R}_{ n n}^{b}(\bm k)\approx e^{-iq_b\mathcal{R}^b_{ n n}(\bm k)}~. 
\ee
Using this into Eq.~\eqref{ScRt2}, we get, 
\be\label{final_SR2}
   w_{n,\kb\kb'}^{(2S)}=\frac{2\pi n_i V_0^2}{\hbar} |\langle u_{ n \bm k}|u_{ n \bm k'}\rangle|^2 \delta(\varepsilon_{ n \bm k} - \varepsilon_{ n \bm k'})\approx \frac{2\pi n_i V_0^2}{\hbar} \delta(\varepsilon_{ n \bm k} - \varepsilon_{ n \bm k'})~.
\ee
Now let's take the case of $w_{ll'}^{(3),A}$. It contains $\langle u_{ n \bm k}|u_{ n \bm k'}\rangle$, $\langle u_{ n \kb''}|u_{ n \kb'}\rangle$ and $\langle u_{ n \kb'}|u_{ n \bm k}\rangle$. We will also assume $q'=(\kb''-\bm{k})\to0$ which leads to $\ket{u_{ n \kb''}}=\ket{u_{ n \bm{k}}}+q_b' \ket{\partial_b u_{ n \bm{k}}}+\frac{1}{2}q_b' q_c'\ket{\partial_b \partial_c u_{ n \bm{k}}}+\cdots$. So the above-mentioned overlaps come out as following
\bea
    \langle u_{ n \bm k}|u_{ n \bm k''}\rangle &=&1-iq_b'\mathcal{R}^b_{ n n}+\frac{1}{2}q_b'q_c'\langle u_{ n \bm k}|\partial_b \partial_c u_{ n \bm k}\rangle~,\nn\\
    \langle u_{ n \bm k''}|u_{ n \bm k'}\rangle &=& 1+i(q_b'-q_b)\mathcal{R}_{ n n}^b +q_b q_c'\langle \partial_c u_{ n \bm{k}}|\partial_b u_{ n \bm{k}}\rangle+\frac{1}{2}q_b q_c \langle u_{ n \bm k}|\partial_b \partial_c u_{ n \bm k}\rangle+\frac{1}{2}q_b' q_c'\langle \partial_b \partial_c u_{ n \bm k}|u_{ n \bm k}\rangle~,\nn\\
    \langle u_{ n \bm k'}|u_{ n \bm k}\rangle &=& 1+iq_b\mathcal{R}^b_{ n n}+\frac{1}{2}q_b q_c\langle \partial_b \partial_c u_{ n \bm k}| u_{ n \bm k}\rangle~. \nn
\eea
Using the above in Eq.~\eqref{sct3}, we can simplify $w_{n,\kb\kb'}^{(3),A}$ as following
\bea\label{w_3a_final}
    w_{n,\kb\kb'}^{(3),A}&=&\frac{4\pi^2 n_i V_1^3}{\hbar}{\rm Im}\sum_{\bm k''}\langle u_{ n \bm k}|u_{ n \bm k''} \rangle\langle u_{ n \bm k''}|u_{ n \bm k'}\rangle \langle u_{ n \bm k'}|u_{ n \bm k}\rangle\delta(\varepsilon_{ n \bm k} - \varepsilon_{ n \bm k'})\delta(\varepsilon_{ n\bm k} - \varepsilon_{ n \bm k''})\nn\\
    &=& \frac{4\pi^2 n_i V_1^3}{\hbar}\sum_{\bm k''}{\rm Im}[1+(q_bq_c+q_b'q_c'-q_bq_c')\mathcal{R}^b_{ n n}\mathcal{R}^c_{ n n}+q_bq_c'\langle \partial_c u_{ n\bm{k}}|\partial_b u_{ n\bm{k}}\rangle\nn\\&&+(q_b q_c+q_b' q_c'){\rm Re}(\langle u_{ n \bm{k}}|\partial_b\partial_c u_{ n \bm{k}}\rangle)]\delta(\varepsilon_{ n \bm k}- \varepsilon_{ n \bm k'})\delta(\varepsilon_{ n\bm k} - \varepsilon_{ n \bm k''})\nn\\
    &=& \frac{4\pi^2 n_i V_1^3}{\hbar}\sum_{ \kb''}q_b q_c'{\rm Im}[\langle \partial_c u_{ n\bm{k}}|\partial_b u_{ n\bm{k}}\rangle]\delta(\varepsilon_{ n \bm k}- \varepsilon_{ n \bm k'})\delta(\varepsilon_{ n\bm k} - \varepsilon_{ n \bm k''})\nn\\
    &=& -\frac{2\pi^2 n_i V_1^3}{\hbar}\sum_{ \kb''}(k_b'-k_b) (k_c''-k_c)\Omega^{bc}_ n(\bm{k}) \delta(\varepsilon_{ n \bm k}- \varepsilon_{ n \bm k'})\delta(\varepsilon_{ n\bm k} - \varepsilon_{ n \bm k''})\nn\\
    &=& \frac{2\pi^2 n_i V_1^3}{\hbar}\sum_{ \kb''}[(k_b'-k_b) (k_c''-k_c)]\epsilon_{bcd}\Omega^{d}_ n(\bm{k}) \delta(\varepsilon_{ n \bm k}- \varepsilon_{ n \bm k'})\delta(\varepsilon_{ n\bm k} - \varepsilon_{ n \bm k''})\nn\\
     &=& \frac{2\pi^2 n_i V_1^3}{\hbar}\sum_{ \kb''}[(\kb'-\bm{k})\times(\kb''-\bm{k})]\cdot\bm{\Omega}_ n(\bm{k}) \delta(\varepsilon_{ n \bm k}- \varepsilon_{ n \bm k'})\delta(\varepsilon_{ n\bm k} - \varepsilon_{ n \kb''})\nn\\
      &=& -\frac{2\pi^2 n_i V_1^3}{\hbar}\sum_{ \kb''}[(\kb''\times\kb')+(\kb'\times\bm{k})+(\bm{k}\times\kb'')]\cdot\bm{\Omega}_ n(\bm{k}) \delta(\varepsilon_{ n \bm k}- \varepsilon_{ n \bm k'})\delta(\varepsilon_{ n\bm k} - \varepsilon_{ n \bm k''})~.
\eea
Here $\Omega_{ n}^{bc}$ is the Berry curvature of the system and defined as $\Omega_{ n}^{bc}=-\Omega_{ n}^{cb}=2\mathrm{Im}[\langle\partial_c u_{n\bm k}|\partial_b u_{n\bm k}\rangle]$ for the band $n$. Similarly one can also simplify $w_{ll'}^{(4),A}$ which contains terms as ${\rm Im}[\langle u_{ n\bm{k}} | u_{ n'\kb'''} \rangle \langle u_{ n'\kb'''} | u_{ n\kb'} \rangle \langle u_{ n\kb'} | u_{ n\kb''} \rangle
\langle u_{ n\kb''} | u_{ n\bm{k}} \rangle]$, ${\rm Im}[\langle u_{ n\bm{k}} | u_{ n\kb''} \rangle \langle u_{ n\kb''} | u_{ n'\kb'''} \rangle\langle u_{ n'\kb'''} | u_{ n\kb'} \rangle\langle u_{ n\kb'} | u_{ n\bm{k}}\rangle]$ and ${\rm Im}[\langle u_{ n\bm{k}} | u_{ n'\kb'''} \rangle \langle u_{ n'\kb'''} | u_{ n\kb''} \rangle \langle u_{ n\kb''} |u_{ n\kb'} \rangle\langle u_{ n\kb'} | u_{ n\bm{k}}\rangle]$. with the same approximations as before along with $q''=(\kb'''-\bm{k})\to 0$, we get 
\bea
{\rm Im}[\langle u_{ n\bm{k}} | u_{ n'\kb'''} \rangle \langle u_{ n'\kb'''} | u_{ n\kb'} \rangle \langle u_{ n\kb'} | u_{ n\kb''} \rangle
\langle u_{ n\kb''} | u_{ n\bm{k}} \rangle]={\rm Im}[q_b''(q_c''-q_c)\mathcal{R}^{b}_{nn'}\mathcal{R}^{c}_{n'n}]&=&-\frac{1}{2}q_b''(q_c''-q_c)\Omega^{bc}_{ n n'}~,\nn\\
{\rm Im}[\langle u_{ n\bm{k}} | u_{ n\kb''} \rangle \langle u_{ n\kb''} | u_{ n'\kb'''} \rangle\langle u_{ n'\kb'''} | u_{ n\kb'} \rangle\langle u_{ n\kb'} | u_{ n\bm{k}}\rangle]={\rm Im}[(q_b''-q_b')(q_c''-q_c)\mathcal{R}^{b}_{nn'}\mathcal{R}^{c}_{n'n}]&=&-\frac{1}{2}(q_b''-q_b')(q_c''-q_c)\Omega^{bc}_{ n n'}~,\nn\\
{\rm Im}[\langle u_{ n\bm{k}} | u_{ n'\kb'''} \rangle \langle u_{ n'\kb'''} | u_{ n\kb''} \rangle \langle u_{ n\kb''} |u_{ n\kb'} \rangle\langle u_{ n\kb'} | u_{ n\bm{k}}\rangle]={\rm Im}[q_b''(q_c''-q_c')\mathcal{R}^{b}_{nn'}\mathcal{R}^c_{n'n}]&=&-\frac{1}{2}q_b''(q_c''-q_c')\Omega^{bc}_{ n n'}~.\nn
\eea
We emphasize that setting $n' = n$renders the product $\mathcal{R}^{b}_{nn'} \mathcal{R}^{c}_{n'n}$ purely real. As a result, its imaginary part—and hence the corresponding intraband contribution to the scattering rate—vanishes identically. Therefore, only interband processes ($n'\neq n$) yield a finite contribution. Substituting this constraint into Eq.~\eqref{sct4}, we obtain
\bea
w^{(4),A}_{n,\kb\kb'}&=&\dfrac{2\pi^2 n_i^2 V_0^4}{\hbar}\sum_{ n'}\sum_{\kb''\kb'''}\bigg[q_b''(q_c''-q_c)-(q_b''-q_b')(q_c''-q_c)-q_b''(q_c''-q_c')\bigg]\Omega^{bc}_{ n n'}\dfrac{\delta(\varepsilon_{ n \bm{k}}-\varepsilon_{ n \kb'})\delta(\varepsilon_{ n\kb'}-\varepsilon_{ n\kb''})}{(\varepsilon_{ n\kb'}-\varepsilon_{ n'\kb'''})}\nn\\
&=& \dfrac{2\pi^2 n_i^2 V_0^4}{\hbar}\sum_{ n'}\sum_{\kb''\kb'''}\epsilon_{bcd}\bigg[(k_b'''-k_b)(k_c''-k_c')-(k_c'''-k_c)(k_b'''-k_b'')\bigg]\Omega^{d}_{ n n'}\dfrac{\delta(\varepsilon_{ n \bm{k}}-\varepsilon_{n \kb'})\delta(\varepsilon_{ n\kb'}-\varepsilon_{ n\kb''})}{(\varepsilon_{ n\kb'}-\varepsilon_{ n'\kb'''})}\nn\\
&=& \dfrac{2\pi^2 n_i^2 V_0^4}{\hbar}\sum_{ n'}\sum_{\kb''\kb'''}\bigg[(\kb'''-\bm{k})\times(\kb''-\kb')-(\kb'''-\bm{k})\times(\kb'''-\kb'')\bigg]\cdot\bm{\Omega}_{ n n'}\dfrac{\delta(\varepsilon_{ n \bm{k}}-\varepsilon_{ n \kb'})\delta(\varepsilon_{ n\kb'}-\varepsilon_{ n\kb''})}{(\varepsilon_{ n\kb'}-\varepsilon_{ n'\kb'''})}\nn\\
&=& -\dfrac{2\pi^2 n_i^2 V_0^4}{\hbar}\sum_{ n'}\sum_{\kb''\kb'''}\bigg[(\kb'\times\bm{k})+(\kb''\times\kb')+(\bm{k}\times\kb'')\bigg]\cdot\bm{\Omega}_{ n n'}\dfrac{\delta(\varepsilon_{ n \bm{k}}-\varepsilon_{n \kb'})\delta(\varepsilon_{ n\kb'}-\varepsilon_{ n\kb''})}{(\varepsilon_{ n\kb'}-\varepsilon_{ n'\kb'''})}~.
\eea
Here, $\Omega^{bc}_{ n n'}=-2{\rm Im}\left[ \mathcal{R}_{ n n'}^{b}\mathcal{R}_{ n' n}^{c}\right]$. Now we can also replace $\sum_{\kb'''}1/(\varepsilon_{ n\kb'}-\varepsilon_{ n'\kb'''})\approx 1/(\varepsilon_{ n\kb'}-\varepsilon_{ n'\bm{k}})$ where we have assumed $\kb'''\to \kb$ and lies within the $\rm 3D$-Brillouin zone . So the final expression for the scattering rate comes out as,
\be\label{w_4a_final}
w^{(4),A}_{n,\kb\kb'}=-\dfrac{2\pi^2 n_i^2 V_0^4 }{\hbar}\sum_{ n'}\sum_{\kb''}\bigg[(\kb'\times\bm{k})+(\kb''\times\kb')+(\bm{k}\times\kb'')\bigg]\cdot\bm{\Omega}_{ n n'}\dfrac{\delta(\varepsilon_{ n \bm{k}}-\varepsilon_{ n \bm {k'}})\delta(\varepsilon_{ n\bm{k}}-\varepsilon_{ n\kb''})}{(\varepsilon_{ n\bm{k}}-\varepsilon_{ n'\bm{k}})}~.
\ee
We now specialize to systems with an even energy dispersion under wave-vector inversion,
\begin{equation}\label{energy_cond}
\varepsilon_n(\kb) = \varepsilon_n(-\kb)~,
\end{equation}
a condition naturally satisfied in the presence of either inversion ($\mathcal{P}$) or time-reversal ($\mathcal{T}$) symmetry. 
Under this symmetry, Brillouin-zone summations over odd functions of momentum vanish. In particular,
\begin{equation}
\sum_{\kb'}\kb'\delta(\varepsilon_{n \bm{k}}-\varepsilon_{n\kb'}) = 0~,
\qquad
\sum_{\kb''}\kb''\delta(\varepsilon_{n \bm{k}}-\varepsilon_{n \kb''}) = 0~.
\end{equation}
Furthermore, using the definition of the band-resolved density of states (DOS),
\begin{equation}
\mathcal{D}_n(\varepsilon)
=
\sum_{\kb'}
\delta(\varepsilon - \varepsilon_{n\kb'})~,
\end{equation}
the remaining momentum summations reduce to
\begin{equation}
\sum_{\kb'} 
\delta(\varepsilon_{n\kb} - \varepsilon_{n\kb'})
=
\mathcal{D}_n(\varepsilon_{n\kb})~.
\end{equation}
Substituting these symmetry constraints into Eq.~\eqref{w_3a_final} and Eq.~\eqref{w_4a_final}, we get the simplified form of Eq.~\eqref{w_3a_comp} and Eq.~\eqref{w_4a_comp}, respectively.
\section{Simplification of the side-jump velocity in terms of Berry curvature}\label{app_C}
Starting from Eqs.~\eqref{sj_vel} and \eqref{scrt2}, the side-jump velocity can be written as
\be 
{\bm v}^{\rm sj}_{n}(\kb) =-  \frac{2\pi n_i V_0^2}{\hbar} \sum_{n'}\sum_{\kb'}|\langle u_{ n \bm k}|u_{ n \bm k'}\rangle|^2 \delta(\varepsilon_{ n \bm k} - \varepsilon_{ n \bm k'}) \delta {\bm r}_{nn}(\kb,\kb')~.
\ee 
Here, $\delta r^a_{nn}(\kb,\kb')$ denotes the coordinate shift during an elastic scattering event and is given by
\be\label{eq:pos_shift_app}
\delta r^a_{nn}(\kb, \kb') = R^a_{nn}(\kb)  - R^a_{nn}(\kb') - (\partial_a + \partial'_a) {\rm arg}(\langle u_{ n \bm k}|u_{ n \bm k'}\rangle)~,
\ee
where the intraband Berry connection is
\begin{equation}
R^a_{nn}(\kb)
= \langle u_{n\kb} | i\partial_a u_{n\kb} \rangle~.
\end{equation}
Now using the identity of Eq.~\eqref{expand_u}, we can show that \( \partial'_a \vert u_{n\kb'} \rangle =  \ket{\partial_a u_{n \kb}} + q_b  \ket{ \partial_a \partial_b u_{n \kb}}\), which allows us to express Berry connection at $\kb'$ as 
\be 
R^a_{nn}(\kb') = R^a_{nn}(\kb)  + i q_b \bigg[ \bra{u_{n\kb}} \partial_a\partial_b u_{n\kb} \rangle  + \bra{\partial_b u_{n\kb}} \partial_a u_{n\kb} \rangle\bigg]~. \nn
\ee 
Employing the identity \( \bra{u_{n\kb}} \partial_a\partial_b u_{n\kb} \rangle = - \bra{\partial_a u_{n\kb}} \partial_b u_{n\kb} \rangle - i \partial_a R^b_{nn} \), we obtain 
\be 
R^a_{nn}(\kb') = R^a_{nn}(\kb) - q_b \Omega^{ab}_n(\kb) +  q_b \partial_a R^b_{nn}(\kb)~, \nn
\ee 
where the Berry curvature is
\begin{equation}
\Omega^{ab}_n(\kb)
= i \left[
\langle \partial_a u_{n\kb} | \partial_b u_{n\kb} \rangle
- \langle \partial_b u_{n\kb} | \partial_a u_{n\kb} \rangle
\right]~.
\end{equation}
Now using Eq.~\eqref{expand_uu}, the overlap yields
\begin{equation}
\mathrm{arg}
\langle u_{n\bm k} | u_{n\bm k'} \rangle
= - q_b R^b_{nn}(\kb)~,
\end{equation}
and
\begin{equation}
(\partial_a + \partial'_a)
\mathrm{arg}
\langle u_{n\bm k} | u_{n\bm k'} \rangle
= - q_b \partial_a R^b_{nn}(\kb)~.
\end{equation}
Substituting these expressions into Eq.~\eqref{eq:pos_shift_app}, the coordinate shift simplifies to
\begin{equation}
\delta r^a_{nn}(\kb,\kb')
= q_b \Omega^{ab}_n(\kb)~.
\end{equation}
Using $\Omega^{ab}_n = \epsilon_{abc} \Omega^c_n$, we obtain the compact vector form
\begin{equation}
\delta {\bm r}_{nn}(\kb,\kb')
= (\kb' - \kb) \times {\bm \Omega}_n(\kb).
\end{equation}
In the long-wavelength limit, $\vert \bra{u_{n \kb}} u_{n\kb'}\rangle\vert^2 \approx 1$ and the side-jump velocity reduces to
\be \label{eq:sj_vel_simp}
{\bm v}^{\rm sj}_n (\kb) = \frac{2\pi n_i V_0^2}{\hbar}  \sum_{\kb'} [ (\kb - \kb') \times {\bm \Omega}_n(\kb) ] \delta(\e_{n\kb} -\e_{n\kb'} )~.
\ee 
Under the symmetry condition of Eq.~\eqref{energy_cond}, i.e., $\varepsilon_n(\kb)=\varepsilon_n(-\kb)$, the side-jump velocity reduce to the compact form given in Eq.~\eqref{v_sj_comp}.
\section{Side-jump correction to Spin Current operator}\label{app_D}
In this appendix, we explicitly calculate the spin side jump velocity $J_{a,l}^{\nu,{\rm sj}}$  defined in Eq.~\eqref{total_cur}. 
The disorder-induced correction to the Bloch state $\ket{\delta^{\rm dis} l}$ gives rise to side-jump contribution. To obtain this correction, we employ the Lippmann–Schwinger equation in the presence of an impurity potential $\hat{V}$, $\ket{l^{\rm dis}}=\ket{l}+(\varepsilon_l-H_0+i\eta)^{-1}\hat{T}\ket{l}$ where $\hat{T}=\hat{V}+\hat{V}(\varepsilon_l-H_0+i\eta)\hat{T}$. The disorder-induced correction to the wave function is then $\ket{\delta^{\rm dis}l}=\ket{l^{\rm dis}}-\ket{l}$. Using this correction, the side-jump contribution to the spin current is given by
\begin{equation}
J_{a,l}^{\nu,{\rm sj}}=\langle2 {\rm Re}\bra{l}\hat{j}^{\nu}_a\ket{\delta^{\rm dis} l}+\bra{\delta^{\rm dis }l}\hat{j}^{\nu}_a\ket{\delta^{\rm dis}l}\rangle_{\rm dis},
\end{equation} 
where $\langle...\rangle_{\rm dis}$ denotes disorder averaging. The lowest-order nonvanishing contribution arises at second order in the impurity potential. Explicitly, these terms read
\bea
\langle2 {\rm Re}\bra{l}\hat{j}^{\nu}_a\ket{\delta^{\rm dis} l}\rangle_{\rm dis} &=& 2 {\rm Re} \sum_{l'l''}\dfrac{\bra{l}\hat{j}^\nu_a\ket{l'}\langle V_{l'l''}V_{l''l}\rangle_{\rm dis}}{(\varepsilon_l-\varepsilon_{l'}+i\eta)(\varepsilon_l-\varepsilon_{l''}+i\eta)}
\nn\\&=&2 {\rm Re} \sum_{ n' n''}\sum_{\kb'\kb''}\dfrac{\bra{ n \bm k}\hat{j}^\nu_a\ket{ n'\bm k'}\langle \bra{ n' \bm k'}V\ket{ n''\bm k''}\bra{ n''\bm k''}V\ket{ n\bm k}\rangle_{\rm dis}}{(\varepsilon_{ n \bm k}-\varepsilon_{ n'\bm k'}+i\eta)(\varepsilon_{ n \bm k}-\varepsilon_{ n''\bm k''}+i\eta)}~, \nn\\
\bra{\delta^{\rm dis }l}\hat{j}^{\nu}_a\ket{\delta^{\rm dis}l}\rangle_{\rm dis} &=& \sum_{l'l''}\dfrac{\bra{l'}\hat{j}^\nu_a\ket{l''}\langle V_{ll'}V_{l''l}\rangle_{\rm dis}}{(\varepsilon_l-\varepsilon_{l'}-i\eta)(\varepsilon_l-\varepsilon_{l''}+i\eta)}=\sum_{ n' n''}\sum_{\kb'\kb''}\dfrac{\bra{ n' \bm k'}\hat{j}^\nu_a\ket{ n''\bm k''}\langle \bra{ n \bm k}V\ket{ n'\bm k'}\bra{ n''\bm k''}V\ket{ n\bm k}\rangle_{\rm dis}}{(\varepsilon_{ n \bm k}-\varepsilon_{ n'\bm k'}-i\eta)(\varepsilon_{ n \bm k}-\varepsilon_{ n''\bm k''}+i\eta)}~. \nn
\eea
Here we have used the notation $V_{ll'}=\langle l|V|l'\rangle=\langle n \bm{k}|V| n'\kb'\rangle$. To simplify the expressions above, we adopt a simplest model for disorder consisting of short-range, randomly distributed  $\delta$-function scatterers,$V(\bm{r})=\sum_i V_i\delta(\bm{r}-\bm{R}_i)$. Within this model, the impurity matrix element between Bloch states takes the form
\bea
   \bra{ n \bm k}V\ket{ n'\bm k'}&=&\sum_i\int d{\bm r}V_i\delta(\bm{r}-\bm{R}_i) e^{i(\bm{k}'-\bm{k})\cdot \bm{r}} \langle u_{ n \bm k}|u_{ n' \bm k'}\rangle = \sum_i V_i e^{i(\bm{k}'-\bm{k})\cdot \bm{R}_i} \langle u_{ n \bm k}|u_{ n' \bm k'}\rangle= V^0_{\kb\kb'}\langle u_{ n \bm k}|u_{ n' \bm k'}\rangle~.\nn\\
\eea
Similarly, the matrix element of the spin-current operator can be evaluated as
\bea
   \bra{ n \bm k}\hat{j}^\nu_a\ket{ n'\bm k'}&=&\frac{1}{2\pi}\int d{\bm r}e^{i(\bm{k}'-\bm{k})\cdot {\bm r}}\langle u_{ n \bm k}|\hat{j}^\nu_a|u_{ n'\bm k'}\rangle=\langle u_{ n \bm{k}}|\hat{j}^\nu_a|u_{ n' \bm{k}} \rangle \delta(\kb'-\bm{k})~.
\eea
Using the above, we get the side-jump correction to the $J^\nu_{a,l}$ as follows,
\bea\label{sj_correct}
J_{a,n}^{\nu,{\rm sj}}(\kb)&=&2 {\rm Re} \sum_{\kb'\kb''}\langle V^0_{\kb'\kb''} V^0_{\kb''\kb}\rangle_{\rm dis} \sum_{ n' n''}\dfrac{\langle u_{ n \bm{k}}|\hat{j}^\nu_a|u_{ n' \bm{k}} \rangle \delta(\kb'-\bm{k})\langle u_{ n'\bm k'}|u_{ n''\bm k''}\rangle \langle u_{ n''\bm k''}|u_{ n\bm k}\rangle}{(\varepsilon_{ n \bm k}-\varepsilon_{ n'\bm k'}+i\eta)(\varepsilon_{ n \bm k}-\varepsilon_{ n''\bm k''}+i\eta)}\nn\\
&&+\sum_{\kb'\kb''}\langle V^0_{\kb\kb'} V^0_{\kb''\kb}\rangle_{\rm dis}\sum_{ n' n''}\dfrac{ \langle u_{ n' \kb'}|\hat{j}^\nu_a|u_{ n'' \kb'} \rangle \delta(\kb''-\kb')\langle u_{ n\bm k}|u_{ n'\bm k'}\rangle \langle u_{ n''\bm k''}|u_{ n\bm k}\rangle}{(\varepsilon_{ n \bm k}-\varepsilon_{ n'\bm k'}-i\eta)(\varepsilon_{ n \bm k}-\varepsilon_{ n''\bm k''}+i\eta)}\nn\\
&=& 2{\rm Re} \sum_{\kb''}\langle V^0_{\kb\kb''} V^0_{\kb''\kb}\rangle_{\rm dis} \sum_{ n' n''}\dfrac{\langle u_{ n \bm{k}}|\hat{j}^\nu_a|u_{ n' \bm{k}} \rangle \langle u_{ n'\bm k}|u_{ n''\bm k''}\rangle \langle u_{ n''\bm k''}|u_{ n\bm k}\rangle}{(\varepsilon_{ n \bm k}-\varepsilon_{ n'\bm k}+i\eta)(\varepsilon_{ n \bm k}-\varepsilon_{ n''\bm k''}+i\eta)}\nn\\
&&+\sum_{\kb''}\langle V^0_{\kb\kb''} V^0_{\kb''\kb}\rangle_{\rm dis}\sum_{ n' n''}\dfrac{ \langle u_{ n' \kb''}|\hat{j}^\nu_a|u_{ n'' \kb''} \rangle \langle u_{ n\bm k}|u_{ n'\bm k''}\rangle \langle u_{ n''\bm k''}|u_{ n\bm k}\rangle}{(\varepsilon_{ n \bm k}-\varepsilon_{ n'\bm k''}-i\eta)(\varepsilon_{ n \bm k}-\varepsilon_{ n''\bm k''}+i\eta)}\nn\\
&=& 2{\rm Re} \sum_{\kb''}\langle V^0_{\kb\kb''} V^0_{\kb''\kb}\rangle_{\rm dis} \sum_{ n' n''}\dfrac{J^\nu_{a, n n'}(\bm k) \langle u_{ n'\bm k}|u_{ n''\bm k''}\rangle \langle u_{ n''\bm k''}|u_{ n\bm k}\rangle}{(\varepsilon_{ n \bm k}-\varepsilon_{ n'\bm k})(\varepsilon_{ n \bm k}-\varepsilon_{ n''\bm k''}+i\eta)}\nn\\
&&+\sum_{\kb''}\langle V^0_{\kb\kb''} V^0_{\kb''\kb}\rangle_{\rm dis}\sum_{ n' n''}\dfrac{J^\nu_{a, n' n''}(\bm k'') \langle u_{ n\bm k}|u_{ n'\bm k''}\rangle \langle u_{ n''\bm k''}|u_{ n\bm k}\rangle}{(\varepsilon_{ n'' \bm k''}-\varepsilon_{ n'\bm k''})}\left[\dfrac{1}{(\varepsilon_{ n\bm k}-\varepsilon_{ n'\bm k''}-i\eta)}-\dfrac{1}{(\varepsilon_{ n\bm k}-\varepsilon_{ n''\bm k''}+i\eta)}\right]\nn\\
&=& 2{\rm Re}\left[\sum_{\kb'}\langle V^0_{\kb\kb'} V^0_{\kb'\kb}\rangle_{\rm dis} \sum_{ n' n''}\left(\dfrac{J^\nu_{a, n n'}(\bm k) \langle u_{ n'\bm k}|u_{ n''\bm k'}\rangle \langle u_{ n''\bm k'}|u_{ n\bm k}\rangle}{(\varepsilon_{ n \bm k}-\varepsilon_{ n'\bm k})(\varepsilon_{ n \bm k}-\varepsilon_{ n''\bm k'}+i\eta)}\right.+ \left.\dfrac{J^\nu_{a, n' n''}(\bm k') \langle u_{ n\bm k}|u_{ n'\bm k'}\rangle \langle u_{ n''\bm k'}|u_{ n\bm k}\rangle}{(\varepsilon_{ n'' \bm k'}-\varepsilon_{ n'\bm k'})(\varepsilon_{ n\bm k}-\varepsilon_{ n'\bm k'}-i\eta)}\right)\right]\nn\\
&=& 2\pi{\rm Im}\sum_{ n'n''}\sum_{ \kb'}\langle V^0_{\kb\kb'} V^0_{\kb'\kb}\rangle_{\rm dis} \left[\dfrac{J^\nu_{a, n  n''}(\bm k) \langle u_{ n''\bm k}|u_{ n'\bm k'}\rangle \langle u_{ n'\bm k'}|u_{ n\bm k}\rangle}{(\varepsilon_{ n \bm k}-\varepsilon_{ n''\bm k})}-\dfrac{J^\nu_{a, n' n''}(\bm k') \langle u_{ n\bm k}|u_{ n'\bm k'}\rangle \langle u_{ n''\bm k'}|u_{ n\bm k}\rangle}{(\varepsilon_{ n' \bm k'}-\varepsilon_{ n'' \bm k'})}\right]\nn\\&\times&\delta(\epsilon_{ n \bm{k}}-\epsilon_{ n' \kb'}) \nn\\
&=& -2\pi {\rm Im}\sum_{n' n''}\sum_{ \kb'}W_{\kb\kb'}\left[\dfrac{J^\nu_{a, n' n''}(\bm k') \langle u_{ n\bm k}|u_{ n'\bm k'}\rangle \langle u_{ n''\bm k'}|u_{ n\bm k}\rangle}{(\varepsilon_{ n' \bm k'}-\varepsilon_{ n'' \bm k'})}  -\dfrac{J^\nu_{a, n  n''}(\bm k) \langle u_{ n''\bm k}|u_{ n'\bm k'}\rangle \langle u_{ n'\bm k'}|u_{ n\bm k}\rangle}{(\varepsilon_{ n \bm k}-\varepsilon_{ n''\bm k})}\right]\nn\\&\times& \delta(\epsilon_{ n \bm{k}}-\epsilon_{ n' \kb'})~.
\eea
In arriving at this expression, we have used the identity $\lim_{\epsilon \to 0} \operatorname{Im}\frac{1}{x + i\eta} = -\pi \delta(x)~$. We have also introduced the disorder-averaged scattering amplitude \( W_{{\bm k}{\bm k}'} = \langle |V^{0}_{{\bm k}{\bm k}'}|^{2} \rangle_{\rm dis} \). 

We now further simplify the side-jump correction by adopting the same approximations used in the scattering-rate analysis. Specifically, we assume long-wavelength scattering, such that $\bm q=\bm{k}'-\bm{k}\to 0$, and restrict to intraband processes ($n'=n$). Under these assumptions, Eq.~\eqref{sj_correct} reduces to,  

\bea
J_{a,n}^{\nu,{\rm sj}}(\kb) &=& -2\pi {\rm Im}\sum_{\kb'}W_{\kb\kb'} \delta(\epsilon_{ n \bm{k}}-\epsilon_{ n \kb'})\sum_{ n''\neq n}\left[\dfrac{J^\nu_{a, n n''}(\bm k') \langle u_{ n\bm k}|u_{ n\bm k'}\rangle \langle u_{ n''\bm k'}|u_{ n\bm k}\rangle}{(\varepsilon_{ n \bm k'}-\varepsilon_{ n'' \bm k'})}  -\dfrac{J^\nu_{a, n  n''}(\bm k) \langle u_{ n''\bm k}|u_{ n\bm k'}\rangle \langle u_{ n\bm k'}|u_{ n\bm k}\rangle}{(\varepsilon_{ n \bm k}-\varepsilon_{ n''\bm k})}\right]~.\nn\\
\eea
From long-range scattering approximation ($\kb'-\bm{k}=q\to 0$ and $\kb''-\bm{k}=q'\to 0$) we get 
\bea
\langle{u_{n\bm{k}}|u_{n\kb'}}\rangle=1-iq_b\mathcal{R}^b_{nn}~,~~~
\langle{u_{n\bm{k}}|u_{n''\kb'}}\rangle=-iq_c\mathcal{R}^c_{nn''}~,~~~\langle{u_{n''\bm{k}}|u_{n\kb'}}\rangle=-iq_c\mathcal{R}^c_{n''n}~.
\eea
Putting them also in the same expression,
\bea
J_{a,n}^{\nu,{\rm sj}}(\kb) &=& -2\pi {\rm Im}\sum_{\kb'}W_{\kb\kb'} \delta(\epsilon_{ n \bm{k}}-\epsilon_{ n \kb'})\sum_{ n''\neq n}\left[\dfrac{J^\nu_{a, n n''}(\bm k') (1-iq_b \mathcal{R}_{nn}^b)(iq_c\mathcal{R}_{n''n}^c)}{(\varepsilon_{ n \bm k'}-\varepsilon_{ n'' \bm k'})} \right.\nn\\&&~~~~~~~~~~~~~~~~~~~~~~~~~~~~~~~~~~~~~~~~~~~~~~~~~~~~~~~~~-\left.\dfrac{J^\nu_{a, n  n''}(\bm k) (-iq_c\mathcal{R}_{n''n}^c)(1+iq_b\mathcal{R}_{nn}^b)}{(\varepsilon_{ n \bm k}-\varepsilon_{ n''\bm k})}\right],\nn\\
&\approx& -2\pi {\rm Im}\sum_{\kb'}W_{\kb\kb'} \delta(\epsilon_{ n \bm{k}}-\epsilon_{ n \kb'})\sum_{ n''\neq n}\left[\dfrac{J^\nu_{a, n n''}(\bm k) (iq_c\mathcal{R}_{n''n}^c)}{(\varepsilon_{ n \bm k}-\varepsilon_{ n'' \bm k})}  -\dfrac{J^\nu_{a, n  n''}(\bm k) (-iq_c\mathcal{R}_{n''n}^c)}{(\varepsilon_{ n \bm k}-\varepsilon_{ n''\bm k})}\right],\nn\\
&=& -4\pi {\rm Im}\sum_{\kb'}W_{\kb\kb'} \delta(\epsilon_{ n \bm{k}}-\epsilon_{ n \kb'})\sum_{ n''\neq n}\left[\dfrac{J^\nu_{a, n n''}(\bm k) (iq_c\mathcal{R}_{n''n}^c)}{(\varepsilon_{ n \bm k}-\varepsilon_{ n'' \bm k})}\right],
\eea
Now, under the even-dispersion condition of Eq.~\eqref{energy_cond}, the above expression further simplifies to the compact form given in Eq.~\eqref{j_sj_comp}.
In deriving the final expression, we have used the definition of the spin Berry curvature $\Omega^{\nu;b}_n=-2\hbar^2\epsilon_{bac}{\rm Im}\sum_{n''}\left[\dfrac{J^\nu_{a, n n''} {v}_{n''n}^c}{(\varepsilon_{ n \bm k}-\varepsilon_{ n'' \bm k})^2}\right]$ together with the velocity matrix element $v^c_{n''n}=i(\varepsilon_{n''\bm{k}}-\varepsilon_{n\bm{k}})\mathcal{R}^c_{n''n}/\hbar$ for $n''\neq n$.

\twocolumngrid
\bibliography{Ref}
\end{document}